\documentclass[11pt,a4paper]{article}
\usepackage{amsmath,amsthm,amssymb,amsfonts,mathtools,bm,bbm,array,float,mathdots,dsfont,mathrsfs}
\usepackage{caption,subcaption,multirow,longtable,lscape,wrapfig}
\usepackage[T1]{fontenc} 
\usepackage{ctable}
\usepackage{notoccite,comment}
\usepackage[toc,page]{appendix}
\usepackage[utf8]{inputenc}
\usepackage[autostyle=true]{csquotes}
\usepackage{jheppub}
\usepackage{tabularx}

\title{\fontsize{15.75}{18}\selectfont Calabi-Yau Four/Five/Six-folds as $\mathbb{P}_\textbf{w}^n$ Hypersurfaces:\\ Machine Learning, Approximation, and Generation}
\author[a]{Edward Hirst,}
\author[a]{Tancredi Schettini Gherardini}

\affiliation[a]{
    Centre for Theoretical Physics, Queen Mary University of London, E1 4NS, UK}

\emailAdd{e.hirst@qmul.ac.uk}
\emailAdd{t.schettinigherardini@qmul.ac.uk}

\preprint{\begin{flushright}
QMUL-PH-23-25
\end{flushright}}
\abstract{
Calabi-Yau four-folds may be constructed as hypersurfaces in weighted projective spaces of complex dimension 5 defined via weight systems of 6 weights. In this work, neural networks were implemented to learn the Calabi-Yau Hodge numbers from the weight systems, where gradient saliency and symbolic regression then inspired a truncation of the Landau-Ginzburg model formula for the Hodge numbers of any dimensional Calabi-Yau constructed in this way. The approximation always provides a tight lower bound, is shown to be dramatically quicker to compute (with computation times reduced by up to four orders of magnitude), and gives remarkably accurate results for systems with large weights. Additionally, complementary datasets of weight systems satisfying the necessary but insufficient conditions for transversality were constructed, including considerations of the interior point, reflexivity, and intradivisibility properties.
Overall producing a classification of this weight system landscape, further confirmed with machine learning methods. Using the knowledge of this classification, and the properties of the presented approximation, a novel dataset of transverse weight systems consisting of 7 weights was generated for a sum of weights $\leq 200$; producing a new database of Calabi-Yau five-folds, with their respective topological properties computed. Further to this an equivalent database of candidate Calabi-Yau six-folds was generated with approximated Hodge numbers.

}

\begin{document}
\maketitle

\numberwithin{equation}{section}
\numberwithin{figure}{section}
\numberwithin{table}{section}

%%%%%%%%%%%%%%%%%%%%%%%%%%%%%%%%%%%%%%%%%%%%%%%%%
\section{Introduction}
Calabi-Yau manifolds have been an epicentre for academic breakthroughs since their conception by the late great Professor Eugenio Calabi \cite{calabi1954}, some 80 years ago.
Amplified by the awarding of a Fields medal for the proof of their existence by Professor Shing-Tung Yau \cite{Yau1978OnTR}, their importance within mathematics and to the mathematical community has since been firmly substantiated.
However, beyond their interest in mathematics, these geometries have received notable acclaim within the physics community as well.
For self-consistency, in superstring theory, the space-time within which we live must be 10-dimensional in nature; to ensure compatibility with the 4-dimensional space-time we observe, the remaining 6-dimensions must form some compact geometry, of which Calabi-Yau manifolds and their orbifolds are the most prudent and popular candidates \cite{Candelas:1985en}.

A selection of the defining features of Calabi-Yau manifolds are what makes them so appropriate for string compactification.
Beyond being compact, their K\"ahler $\mathrm{SU(n)}$ holonomy allows them to support the appropriate fields, as fluxes, which can reduce to those seen in the standard model. Moreover, being Ricci-flat in nature, they manifestly satisfy the vacuum Einstein equations desired to incorporate gravity.
Under dimensional reduction of a string theory via Calabi-Yau compactification, many properties of the subsequent 4-dimensional theory become directly dependent on the used Calabi-Yau's geometry, and thus choosing the correct Calabi-Yau becomes paramount to producing a theory that well models the universe.

Unfortunately, the landscape of these geometries is enormous, and its structure largely unknown \cite{He:2018jtw}.
Through a variety of construction methods, billions of these geometries have so far been enumerated \cite{Kreuzer:2000xy,Altman:2014bfa}, and with numbers at this scale brute-force analysis of the corresponding theories becomes computationally infeasible \cite{Demirtas:2020dbm}.
Databases of this size hence require statistical methods of analysis to extract meaningful insight, and, inspired by a multitude of successes in other fields, academics have been recently experimenting with the application of techniques from \textit{machine learning}.

Machine learning is a broadly-used umbrella term for techniques in computational statistics; loosely separated into 3 subfields: supervised, unsupervised, and reinforcement learning \cite{Ruehle:2020jrk,Bao:2022rup,He:2023csq}.
The first subfield of supervised learning can be considered as advanced techniques in function fitting, requiring both input and output data to fit. 
The second of unsupervised learning includes more general feature analysis and dimensional reduction, looking at input data on its own.
The final subfield is reinforcement learning, which trains an agent to search a space of potential solutions for an optimum.

With use notably initiated for the string community in the simultaneous works \cite{He:2017aed,Krefl:2017yox,Ruehle:2017mzq,Carifio:2017bov}, successes inspired applications in a broader range of contexts.
Beyond many excellent programs of work seeking to numerically construct the elusive Ricci-flat Calabi-Yau metrics \cite{Headrick:2005ch,Douglas:2006rr,Ashmore:2019wzb,Anderson:2020hux,Douglas:2020hpv,Jejjala:2020wcc,Larfors:2021pbb,Ashmore:2021ohf,Larfors:2022nep,Berglund:2022gvm,Gerdes:2022nzr} enabling further steps in string phenomenology \cite{Ashmore:2023ajy,Ahmed:2023cnw}, there has been a variety of papers finding real efficacy of machine learning in predicting Calabi-Yau topological properties.

In particular, supervised methods have been especially amenable to the prediction of Hodge numbers, where expensive and difficult computations can be avoided if statistically confident predictions indicate a candidate geometry is highly unlikely to be relevant for one's desired application.
This has been shown in the Calabi-Yau \textit{three-fold} (i.e. 3 complex dimensional) construction cases of weighted projective spaces \cite{He:2017aed,Berman_2022}, complete intersections \cite{He:2017aed,Bull:2018uow,Bull:2019cij,Brodie:2019dfx,Erbin:2020srm,aslan2023group,Erbin_2023}, their generalised cases \cite{Cui:2022cxe}, and via toric varieties \cite{Klaewer:2018sfl,Berglund:2021ztg}.

Whilst Calabi-Yau three-folds are excellent candidates for superstring compactification from 10-dimensions, superstring theory also has interpretations within its parent theories of M-theory and F-theory, which are 11 and 12-dimensional, respectively.
Therefore, to compactify these higher-dimensional theories down to 4-dimensions, higher-dimensional geometries are needed.
M-theory compactification requires 7-dimensional manifolds \cite{Gherardini:2023uyx}, notably G$2$-manifolds exhibiting evermore elusive constructions, machine learning in this area has been initiated by recent work considering the related G$2$-structure geometries with success predicting their equivalent Hodge numbers \cite{Aggarwal:2023swe}.
Alternatively, F-theory compactification requires Calabi-Yau \textit{four-folds}, where machine learning methods have been effective for the complete intersection construction\footnote{We note recent successes in generating and machine learning Calabi-Yau five-folds \cite{https://doi.org/10.1002/prop.202300262}, where the first systematic construction of complete intersection Calabi-Yau five-folds is presented, producing a large but inexhaustive list of new spaces.} \cite{He:2020lbz,Erbin:2021hmx}, for which an exhaustive list has been determined \cite{Gray_2013, Gray_2014}; however, machine learning methods have not yet been tested for the other constructions.

Whilst the database of Calabi-Yau four-folds from weighted projective spaces has been constructed  \cite{Kreuzer:1997zg,Lynker_1999,Brown_2015}, the toric variety construction method is too large to be enumerated in full \cite{Scholler:2018apc}, despite new work showing machine learning methods can help search this intractable space \cite{Berglund:2023ztk}.
Therefore, inspired by an array of successes in machine learning Calabi-Yau three-folds, this work looks to examine the suitability of these methods to the yet untouched database of Calabi-Yau four-folds built from weighted projective spaces, whilst developing on techniques inaugurated in \cite{Berman_2022}.

This paper begins by detailing the Calabi-Yau construction of interest in §\ref{sec:background}, followed by analysis of the weight system data and respective topological invariants in §\ref{sec:dataanalysis}, with detail on generated complementary datasets.
In §\ref{sec:ml}, the machine learning methods used are introduced, followed by their application, results, and interpretation.
Central to this work, in §\ref{sec:Approximation}, is the presentation of an approximation formula for computation of Hodge numbers of Calabi-Yau manifolds constructed via weighted projective spaces, providing a tight lower bound, and enormous improvements in computation time.
In §\ref{sec:higherweights} this approximation, as well as the other related properties, is used to construct candidate transverse weight systems of 7 and 8 weights with sum of weights up to 200, along with the topological properties of the subsequent Calabi-Yau five- and six-folds respectively.
Finally, in §\ref{sec:conclusion}, results are summarised and outlook applications discussed.

The code for this work was completed in \texttt{python}, with use of machine learning libraries \texttt{scikit-learn} \cite{scikit-learn} and \texttt{tensorflow} \cite{tensorflow2015-whitepaper}; datasets and scripts are made available at this paper's respective repository on \href{https://github.com/Tancredi-Schettini-Gherardini/P5CY4ML}{GitHub}\footnote{\url{https://github.com/Tancredi-Schettini-Gherardini/P5CY4ML}}.

As a final comment, we leave some reference to exciting applications of machine learning across alternative subfields in mathematical physics, which have included work on amoeba \cite{Bao:2021olg,Chen:2022jwd,Seong:2023njx}, branes \cite{Halverson:2019tkf,Loges:2021hvn,Arias-Tamargo:2022qgb,Loges:2022mao}, conformal theories \cite{Chen:2020dxg,Kantor:2021kbx,Kantor:2021jpz,Kantor:2022epi,Niarchos:2023lot}, quivers \cite{Bao:2020nbi,Dechant:2022ccf,Cheung:2022itk,Chen:2023whk}, phenomenology \cite{Abel:2014xta,Bies:2020gvf,Krippendorf:2021uxu,Constantin:2021for,Abel:2021rrj,Berman:2022jqn,Abel:2023zwg,Dubey:2023dvu}, and other related geometry \cite{He:2020eva,Bao:2021auj,Bao:2021ofk,Gao:2021xbs,coates2022machine,Coates2023,coates2023machine,Manko:2022zfz,Choi:2023rqg}.
With an abundance of mathematical objects used throughout physics, the age of application of machine learning to uncover new physical understanding is, alluring, just at its beginning.

%%%%%%%%%%%%%%%%%%%%%%%%%%%%%%%%%%%%%%%%%%%%%%%%%
\section{Background}\label{sec:background}

The most natural appearance of Calabi-Yau four-folds is in the context of $N=1$ compactification of F-theory to four dimensions, which was studied in the seminal works of \cite{1996,Klemm_1998,Gukov:1999ya}, among others. 

F-theory emerges upon geometrisation of the axio-dilaton present in Type IIB superstring theory, resulting in a $12$-dimensional theory. It was first developed in the seminal work of Vafa \cite{1996}.
A Calabi-Yau four-fold, which is elliptically fibred, serves as the internal space for the compactification of F-theory to an $N=1$ supersymmetric theory in four dimensions (see \cite{donagi2012model}, for instance). As usual, the moduli space is determined by the possible deformation families, encoded in the cohomological data, which are the main subject of this work. Moreover, Calabi-Yau four-folds can also appear in the compactification of M-theory to three dimensions, leading to an $N=2$ supersymmetric theory. The two reductions are linked when the four-fold $X$ is elliptically fibred, as shown in \cite{Klemm_1998}. For a given fibration $\mathcal{E} \xrightarrow{} X \xrightarrow[]{\pi} B$, a compactification of M-theory on $X$ coincides with a compactification of F-theory on $X \times S^1$. The set of Calabi-Yau four-folds studied in this paper also includes spaces with negative Euler number, a feature that allows for supersymmetry breaking in M-theory compactification. 

\paragraph{The Construction}\mbox{}\\
The Calabi-Yau four-folds considered in this work are constructed as codimension-1 hypersurfaces in compact complex 5-dimensional weighted projective spaces $\mathbb{P}^5_\textbf{w}$.
A general $n$-dimensional weighted projective space $\mathbb{P}_\textbf{w}^n$ is defined by considering $\mathbb{C}^{n+1}$, spanned by coordinates $\{ z_0 , \cdots , z_{n} \}$, removing the origin to form $\mathbb{C}^{n+1}/\{0\} = (\mathbb{C}/\{0\})^{n+1} = (\mathbb{C}^\ast)^{n+1}$, then subjecting it to an identification given by:
\begin{equation}\label{eq:wps}
    ( z_0 , ..., z_{n} ) \sim ( \lambda^{w_0} z_0 , \cdots , \lambda^{w_n} z_{n} ) \; ,
\end{equation}
for all non-zero complex numbers $\lambda \in \mathbb{C}^\ast$. 
The integer numbers $w_i$'s are called \textit{weights} (hence the name of the construction), and the vector of weights $(w_0,w_1,...,w_n)$ is known as a \textit{weight system} of $n+1$ weights. 
For weight systems to uniquely define weighted projective spaces, the set of weights needs to be coprime, removing redundancy introduced by re-scaling of the identification parameter $\lambda$.
There are infinitely many coprime weight systems which thus each uniquely define a weighted projective space.
However not all of these weighted projective spaces will admit Calabi-Yau hypersurfaces, in the cases they do the weight system is defined as \textit{transverse}.
For transverse weight systems these Calabi-Yau hypersurfaces are then homogeneous functions of specific degree, as required to satisfy the defining vanishing first Chern class property necessary to produce a Calabi-Yau.

To briefly introduce this, start by assuming we have a transverse weight system such that the hypersurface can avoid the singularities of the ambient $\mathbb{P}^n_\textbf{w}$.
This defining hypersurface equation $p=0$ therefore has no common solutions with its derivative $dp=0$. %ie \frac{\partial p}{\partial z_i} = 0 \forall i only if z_i=0 \forall i
Defining $\mathcal{T}_\mathbb{P}$ as the tangent bundle of the ambient $\mathbb{P}^n_\textbf{w}$, such that the hypersurface submanifold $\mathcal{M}$ has respective normal bundle $\mathcal{N}$, then $\mathcal{T}_\mathbb{P} = \mathcal{T}_\mathcal{M} \oplus \mathcal{N}$ allowing the computation of the Chern polynomial for the hypersurface submanifold from that of the ambient space and normal bundle \cite{CANDELAS1990383}.
A tangent space in $\mathcal{T}_\mathbb{P}$ is a space of vectors $v = v^i\frac{\partial}{\partial z^i}$ which act on functions in the ambient space (i.e. functions of the ambient space's homogeneous coordinates $z^i$).
The homogeneous nature of the functions however leads to an identification in this space of vectors $v^i \sim v^i + w_iz^i$, since $\sum_i w_i z^i \frac{\partial}{\partial z^i} f = mf$ for generic homogeneous function $f$ of degree $m$, reducing the space dimension by 1 as required.
The independence of the vectors (except for this identification) leads to a decomposition of the tangent bundle into line bundles $\mathcal{T}_\mathbb{P} = (\mathcal{O}(w_0) \oplus \mathcal{O}(w_1) \oplus ... \oplus \mathcal{O}(w_n))/\mathcal{O}$, with the trivial bundle in the denominator.
The Chern polynomial of these 1-dimensional line bundles is then $c(\mathcal{O}(w_i)) = 1+w_i\omega$, for $\omega$ the K\"ahler form of the ambient space, leading to
\begin{equation}
    c(\mathcal{T}_\mathbb{P}) = \Pi_i (1+w_i\omega)\;.
\end{equation}
Whereas, since the degree $\mathfrak{d}$ hypersurface equation is codimension 1, it can be viewed as a fibre coordinate for $\mathcal{N}$, such that $\mathcal{N}=\mathcal{O}(\mathfrak{d})$, and hence $c(\mathcal{O}(\mathfrak{d})) = 1+\mathfrak{d}\omega$.
Therefore the overall Chern polynomial for the hypersurface submanifold tangent space is
\begin{equation}
    c(\mathcal{T}_\mathcal{M}) = \frac{\Pi_i (1+w_i\omega)}{1+\mathfrak{d}\omega}\;,
\end{equation}
where the first Chern class, $c_1$, can be extracted by expansion of the above, leading to the condition: $c_1(\mathcal{T}_\mathcal{M}) = (\sum_i (w_i) -\mathfrak{d})\text{Tr}(\omega)$, which for the hypersurface to define a Calabi-Yau manifold requires $c_1=0$, causing
\begin{equation}
    \mathfrak{d} = \sum_i w_i\;.
\end{equation}
Therefore Calabi-Yau manifolds can be constructed as hypersurfaces in weighted projective spaces, where the weights form a transverse weight system and the hypersurface is defined by a homogeneous equation of degree equal to the sum of the weight system weights, $w_{tot} \vcentcolon = \sum_i w_i$.

\paragraph{Weight System Properties}\mbox{}\\
Consequently, for the central focus of this work, which is Calabi-Yau four-folds, we are interested in weight systems consisting of 6 weights.
For the hypersurface to be Calabi-Yau, the weight system must be transverse, which synonymously in the mathematics literature may also be referred to as quasi-smooth, in that the hypersurface has no additional singularities other than those inherited from the ambient space.
The complete list of transverse weight systems of 6 weights was classified in \cite{Lynker_1999}, totalling 1100055.
Transverse weight systems were first bulk-generated in \cite{CANDELAS1990383} for the three-fold case of 5-weight weight systems, later extended to the full finite list of 7555 in  \cite{Klemm:1992bx,Kreuzer:1992np}. In general, the number of transverse weight systems was proved to be finite for any dimension in \cite{Kreuzer:1992bi}.
These constructions, as well as those for four-folds in \cite{Lynker_1999}, relied heavily on the use of these hypersurfaces as potentials in Landau-Ginzburg theories \cite{Vafa:1988uu,Witten:1993yc}, with limited direct interpretation in terms of the weights.
However, in the original construction in \cite{CANDELAS1990383}, a necessary but insufficient condition for transversality was introduced in terms of the weights exclusively.

This necessary but insufficient condition for a general dimensional weight system to be transverse is based on divisibility between these weights.
We thus dub this property \textit{intradivisibility}. A weight system is intradivisible iff 
\begin{equation}\label{eq:intradivisibility}
    \forall w_i \ \exists w_j \ s.t. \ \frac{w_{tot} - w_j}{w_i} \in \mathbb{Z}^+\;,
\end{equation}
such that each weight can be subtracted from the sum of the weights and the result will be divisible by a weight in the weight system\footnote{We note a nomenclature subtlety in \cite{Berman_2022} where `transverse' was used to depict a weight system satisfying this property, and `Calabi-Yau' was used to depict a weight system which can admit a Calabi-Yau hypersurface. In this work, we reserve `transverse' for the weight systems with Calabi-Yau hypersurfaces where the solutions to the hypersurface equation and its derivative are transverse, and introduce `intradivisibility' for weight systems satisfying the property of \eqref{eq:intradivisibility}}.
This property can be computed from the weights alone, and provides a means of identifying weight systems which are certainly not transverse -- where this condition does not hold.

In addition to intradivisibility, another property of a weight system is required for it to be transverse, and this property comes from the more general toric interpretation of the weighted projective spaces\footnote{These can be alternatively generalised to fake weighted projective spaces \cite{Kasprzyk_2009}, but this is another story.}.
In this interpretation the ambient $\mathbb{P}^n_\textbf{w}$ are toric varieties, defined by fans in $\mathbb{R}^n$, which can be built from convex lattice polytopes in $\mathbb{Z}^n$ centred on the origin.

A polytope \cite{Bao:2021ofk} is itself defined by a collection of $d$ hyperplane inequalities such that all points $x \in \mathbb{R}^n$ are in the polytope if $H \cdot x \geq b$ for some defining $d \times n$ matrix $H$ and constant $d$-vector $b$.
The constituent parts of the polytope as defined by the intersection of the hyperplanes are the polytope faces, where 0-faces are vertices, 1-faces are edges, and so on up to $(n-1)$-faces which are facets.
In the case where the vertices' coordinates are all integers across the polytope, the polytope is a lattice polytope\footnote{Lattice polytopes can also be physically interpreted as toric diagrams of quiver gauge theories \cite{Hanany:2005ve}.}.
If the polytope contains a single lattice point in its strict interior, then the polytope is called \textit{interior point}, or \textit{IP}; due to the affine symmetry of the lattice this point can always be shifted to be the origin\footnote{In the mathematics literature lattice polytopes with exclusively the origin in the strict interior are called \textit{Fano}, since they lead to Fano varieties. Where the boundary lattice points are only the polytope's vertices the polytope is terminal Fano, whilst where there're extra boundary lattice points the polytope is canonical Fano \cite{Kasprzyk_2010}.}. %both these terminal and canonical singularites can be resolved
The respective toric fan for a lattice polytope is defined by constructing 1-cones which are lines connecting the origin to each vertex, then extending each line infinitely.
The remaining higher cones of the fan are then defined by the intersections of the polytope hyperplanes\footnote{To ensure a smooth toric variety the polytope can first be triangulated to resolve the singularities arising from the interior points of the facets.}. %note want a star triangulation to construct the fan, and physcially care about fine and regular also
The toric variety \cite{cox2011toric} is then constructed from the fan through consideration of its respective dual fan, each cone in the fan has a dual cone which is the set of all points whose inner product with points in the cone produces a non-negative number.
The union of all dual cones is the dual fan.
Finally, the toric variety is defined as the maximal spectrum of the generators of the dual fan's 1-cones, i.e. taking the dual fan 1-cone generators (vectors in $\mathbb{Z}^n$) and treating their entries as exponents of the coordinates in some $\mathbb{C}^n$, each generator provides a condition on the coordinate ring $\mathbb{C}^n$, and the resulting spectrum of maximal ideals of this quotient ring defines the toric variety. %here assuming both lattice and dual are \mathbb{Z}^n, also avoiding mention of schemes

From this construction it has been shown how polytopes lead to toric varieties, despite the series of steps needed to go from the polytope to the variety, a surprising amount about the variety can be deduced from the polytope information alone.
An example of this is that for the variety to be compact, the dimension of the polytope must equal the dimension of the lattice it is defined on. %variety is gorenstein if all weights divide the sum of weights \cite{Batyrev:2020vnd}
In fact, in this vein there is a more direct construction method for the toric variety from the polytope, and one more similar to the weighted projective space construction.
Whereas, where weighted projective spaces are defined through one identification of $\mathbb{C}^{n+1}$ using one weight system as in \eqref{eq:wps}, this can be generalised to $k$ identifications of $\mathbb{C}^{n+k}$ using $k$ weight systems.
If these weights are selected to be vectors spanning the kernel of the lattice polytope's vertex matrix\footnote{We note that the $GL(k,\mathbb{Z})$ symmetry in the kernel basis manifests itself as the redundancy from the linear addition of weight systems before identification. Additionally, on more general lattices there are further identifications from the lattice gradings to account for where the vertices are not primitive.} then the generated variety is the same toric variety as that constructed via the dual fan method above. %a primitive vertex has gcd of its coordinates equal to 1

In this way, the weight systems of consideration for weighted projective spaces can be generalised to include \textit{combined weight systems} (with many weight systems) for toric varieties.
The hypersurface equation defining the potential Calabi-Yau in the weighted projective space then becomes a generic hypersurface in the toric variety's anticanonical divisor class \cite{Batyrev:1993oya}; with alternative interpretations as non-transverse hypersurfaces in weighted projective spaces \cite{Candelas:1994bu}. %in dim 3 the CY automatically smooth, in dim 4 can have gorenstein point-like singularities which can be avoided by the hypersurface, in higher dim the singularity structure and resolution is more subtle
Inverting the process of extracting weights from a polytope, polytopes can also be constructed from (combined) weight systems.
However, before introducing this, the definition of a polytope's dual is needed.

In a similar way to how a polytope's fan has a dual fan, a polytope has a dual polytope defined as the set of points such that the inner product between any point in the polytope and any point in the dual polytope $\geq -1$.
By definition, the dual of an IP polytope is hence also IP \cite{Bouchard:2007ik}.
However the dual of lattice polytope is not necessarily lattice, and in the special cases where both a polytope and its dual are lattice the polytopes are denoted as a reflexive pair -- both satisfying the reflexivity property. 
In fact, it is one of the astounding beauties of the toric construction of Calabi-Yau's that the hypersurfaces in toric varieties from dual lattice polytopes are in fact mirror symmetric \cite{Batyrev:1993oya}.
A weight system is thus \textit{reflexive} if the lattice polytope constructed from it is reflexive.

Let us return to the construction of an IP polytope from an IP weight system.
Here, the hyperplane equations defining the polytope include an equality for each weight system such that $\sum_i w_i x_i = w_{tot}$, and inequalities defined by $x_i \geq 0 \ \forall i$ \cite{Batyrev:2020vnd}. %can affine transform this polytope to make other formulations of the equalities / inequalities equivalent
Through this construction, the point $x_i = 1 \ \forall i$ is manifestly contained within the polytope, since it naturally satisfies the equalities and inequalities; noting that an affine transformation can set this point to be the origin. 
This general polytope is hence always IP, however it may be rational and we are often more interested in its restriction to a lattice. %usually integer lattice

To be able to then define and check the IP property of a given weight system it suffices to construct the respective polytope, and consider it as existing on the crudest lattice it can (that generated by the polytopes vertices). 
The dual polytope can then be generated from this, and respectively the dual lattice (all real points which dot product with all points in the polytope's lattice to integers), however the vertices of the dual polytope may not lie on the dual lattice, and thus a restriction is required by taking the convex hull of dual lattice points that lie within/on this dual polytope.
This restriction may slice parts of the dual polytope off, producing a smaller dual polytope, which when taking the dual again will produce an new version of the original polytope, which we define to be the integer polytope of interest, and in doing this the new boundaries may intersect the origin.
Therefore the new restricted polytope may no longer contain the origin in its interior, and would thus not be IP \cite{Scholler:2018apc}.
In the cases where the origin does remains in the strict interior the respective lattice polytope is IP, and we define the weight system to be IP too. %Defn 1.2 in Batyrev:2020vnd must be wrong? 

All real polytopes constructed from weight systems are simplices, since there are as many intersections of the single defining equality with the inequalities as there're lattice dimensions (and also weights); equivalently, those from the larger combined weight systems are the union of simplices.
However the restriction to the relevant lattice as described above may generalise the polytope causing the lattice polytopes to be unions of simplices also.
For the weight system to be transverse there must be no further unavoidable singularities than the origin, and this translates to having no interior points on the polytope facets. %need to triangulate the polytope first
It is where this occurs that weight systems can be IP but not transverse. %probably slightly more general than this

The importance of the IP property for weight systems comes from \cite{Skarke:1996hq}, where it was shown that any transverse weight system is by \textit{necessity} IP for any size weight system.
However the converse is not true, and thus overall we have 2 independent necessary but insufficient conditions for a weight system to be transverse: intradivisibility and IP.
Beyond this we have another weight system property: reflexivity; where its interrelation with transversality depends on the construction dimension in question \cite{Kreuzer:1997zg,Batyrev:2020vnd}.
Denoting the sets of IP, reflexive, and transverse weight systems of $n$ weights by $IP(n)$, $R(n)$, and $T(n)$ respectively, with their respective sizes as $|IP(n)|$, $|R(n)|$, and $|T(n)|$, the relations between, and frequencies of, weight systems with each property are shown in Table \ref{tab:WS_properties}.

\begin{table}[tb]
\centering
\begin{tabular}{|c|c|ccc|}
\hline
\multirow{2}{*}{\begin{tabular}[c]{@{}c@{}}Number of\\ weights\end{tabular}} & \multirow{2}{*}{Relations}                                                                              & \multicolumn{3}{c|}{Property}                                                    \\ \cline{3-5} 
                                                                             &                                                                                                         & \multicolumn{1}{c|}{$|IP(n)|$}    & \multicolumn{1}{c|}{$|R(n)|$}     & $|T(n)|$ \\ \hline
2                                                                            & $IP(2) = R(2)$                                                                                          & \multicolumn{1}{c|}{1}            & \multicolumn{1}{c|}{1}            & -        \\ \hline
3                                                                            & $IP(3) = R(3) = T(3)$                                                                                   & \multicolumn{1}{c|}{3}            & \multicolumn{1}{c|}{3}            & 3        \\ \hline
4                                                                            & $IP(4) = R(4) = T(4)$                                                                                   & \multicolumn{1}{c|}{95}           & \multicolumn{1}{c|}{95}           & 95       \\ \hline
5                                                                            & $IP(5) = R(5) \supset T(5)$                                                                             & \multicolumn{1}{c|}{184026}       & \multicolumn{1}{c|}{184026}       & 7555     \\ \hline
6                                                                            & \begin{tabular}[c]{@{}c@{}}$IP(6) \supset R(6)$\\ $IP(6) \supset T(6)$\end{tabular}                     & \multicolumn{1}{c|}{322383760930} & \multicolumn{1}{c|}{185269499015} & 1100055  \\ \hline
$\geq 7$                                                                     & \begin{tabular}[c]{@{}c@{}}$IP(\geq 7) \supset R(\geq 7)$\\ $IP(\geq 7) \supset T(\geq 7)$\end{tabular} & \multicolumn{1}{c|}{?}            & \multicolumn{1}{c|}{?}            & ?        \\ \hline
\end{tabular}
\caption{Known relations between the sets of weight systems satisfying the IP ($IP(n)$), reflexive ($R(n)$), and transverse ($T(n)$) properties as the number of weights in the weight system increases. Additionally showing the sizes of the sets of weight systems satisfying those properties (denoted with $|\cdot |$) at each weight system size. The weight systems for dimensions $\geq 7$ have not been fully computed yet.}
\label{tab:WS_properties}
\end{table}

Due to the need for an identification, weight systems are not defined for 1 weight, and since the transverse property requires taking a codimension-1 hypersurface this property is also not defined for weight systems of 2 weights.
For 2 weights the single weight system is (1,1), which is equivalent to the single 1-dimensional IP and reflexive polytope with vertices a distance 1 either side of the origin.
For 3 weights there are 3 IP weight systems $\{(1,1,1),(1,1,2),(1,2,3)\}$ which are all both reflexive and transverse, corresponding to 3 of the 5 reflexive triangles. %others related to combined weight systems and subpolytopes
Stepping to 4 weights the number grows to 95  \cite{Skarke:1996hq}.
Whilst for 5 weights there is no longer an equality between all these sets of weights, with set sizes computed in \cite{Skarke:1996hq,Klemm:1992bx,Kreuzer:1992np}.
The weight systems of central focus in this work have 6 weights, and it is at this stage that each set of weight systems becomes distinct, where the $IP(6)$ and $R(6)$ sets were computed in \cite{Scholler:2018apc}, and the $T(6)$ set in \cite{Lynker_1999}. %note still overlap as specified by the relation
Beyond systems with 6 weights the set sizes are unknown, as constructions have not yet been attempted (until this work as detailed in §\ref{sec:higherweights}).
The intradivisibility property is not believed to have a finiteness bound, which is why it is not included in these count considerations.
The interrelation of this property with the others is discussed in more detail in §\ref{sec:Partitioning}.
%add explicit examples of weight systems satisfying each property?

\paragraph{Topological Properties}\mbox{}\\
As previously mentioned, the cohomological data of the Calabi-Yau used in compactification determines the moduli space of the resulting compactified supersymmetric theory.
Since Calabi-Yau manifolds are manifestly complex and K\"ahler \cite{Bouchard:2007ik}, the complexity allows use of the decomposition of the complexified cotangent bundle into holomorphic and antiholomorphic parts via eigenspaces of the complex structure: $\mathcal{T}_\mathcal{M}^\ast = \mathcal{T}^{\ast \ (1,0)}_\mathcal{M} \oplus \mathcal{T}^{\ast \ (0,1)}_\mathcal{M}$.
This in turn causes a decomposition of the differential forms as exterior products of these holomorphic and antiholomorphic cotangent bundles $\Lambda^{k}\mathcal{T}_\mathcal{M}^\ast = \bigoplus_{p+q=k} \Lambda^p \mathcal{T}^{\ast \ (1,0)}_\mathcal{M} \otimes \Lambda^q \mathcal{T}^{\ast \ (0,1)}_\mathcal{M}$, such that the $(p,q)$-forms are sections of each sum component with vector space $\Omega^{p,q}(\mathcal{M})$.
From here the cohomology arises using the decomposition of the exterior derivative operator $\text{d} = \partial + \overline{\partial}$, allowing definition of the Dolbeault cohomology groups
\begin{equation}
    H_{\overline{\partial}}^{p,q}(\mathcal{M}) = \frac{Ker(\overline{\partial}: \Omega^{p,q}(\mathcal{M}) \mapsto \Omega^{p,q+1}(\mathcal{M}))}{Im(\overline{\partial}: \Omega^{p,q-1}(\mathcal{M}) \mapsto \Omega^{p,q}(\mathcal{M})))}\;. %kernel and image
\end{equation}
These are defined for all $p$ (arbitrarily they may instead be defined for all $q$ using $\partial$), and the dimension of these groups defines the \textit{Hodge numbers}
\begin{equation}
    h^{p,q} = \dim H_{\overline{\partial}}^{p,q}(\mathcal{M}) \;.
\end{equation}
These Hodge numbers may be arranged into a Hodge diamond, where the symmetries of complex conjugation ($h^{p,q} = h^{q,p}$) and Serre duality ($h^{p,q} = h^{n-p,n-q}$ for $\dim_\mathbb{C}\mathcal{M} = n$) become clear.
\begin{comment}
\begin{equation}
\begin{gathered}
h^{0,0}\\
h^{1,0} \qquad h^{0,1} \\
h^{2,0} \qquad h^{1,1} \qquad h^{0,2} \\
\iddots \ \qquad \iddots \ \qquad \ \ddots \qquad \ \ddots \\
h^{n,0} \qquad \cdots \qquad \cdots \qquad \cdots \qquad h^{0,n} \\
\ddots \ \qquad \ddots \ \qquad \ \iddots \qquad \ \iddots \\
h^{n,n-2} \quad h^{n-1,n-1} \quad h^{n-2,n} \\
h^{n,n-1} \quad h^{n-1,n} \\
h^{n,n}\\
\end{gathered}
\end{equation}
\end{comment}
The K\"ahlerity of the Calabi-Yau manifolds relates these Hodge numbers to the complexified de Rham real cohomological Betti numbers, $b_k$, since
\begin{equation}
\begin{split}
    H_{\text{d}}^{k}(\mathcal{M}) & = \bigoplus_{p+q=k} H_{\overline{\partial}}^{p,q}(\mathcal{M}) \;,\\
    \implies b_k & = \sum_{p+q=k} h^{p,q}\;,
\end{split}
\end{equation}
then also allowing for the manifolds' Euler number, $\chi$, to be computed from these Hodge numbers via $\chi = \sum_k (-1)^k b_k$.

Furthermore, for the specialised K\"ahler case of Calabi-Yau manifolds, there are even further restrictions on these Hodge numbers.
One of the defining properties of a Calabi-Yau manifold is a unique holomorphic top form, which then sets $h^{n,0}=1$, hence also setting to 1 the other Hodge diamond corners (via the conjugation and duality) \cite{Bao:2020sqg}.
Additionally, as the Calabi-Yau's are simply connected, they have trivial first fundamental group and therefore also trivial first homology group, setting $h^{1,0}=0$ and respectively the remaining boundary components of the Hodge diamond \cite{h1994calabi}.
The final Hodge diamond therefore takes the form:
\begin{equation}
\begin{gathered}
1\\
0 \qquad 0 \\
0 \qquad h^{1,1} \qquad 0 \\
0 \qquad h^{1,2} \qquad h^{1,2} \qquad 0 \\
\iddots \ \ \qquad \iddots \qquad\qquad \ddots \qquad \ \ \ddots \\
1 \quad\qquad \cdots \quad\qquad \cdots \quad\qquad \cdots \quad\qquad 1 \\
\ddots \ \ \qquad \ddots \qquad\qquad \iddots \qquad \ \ \iddots \\
0 \qquad h^{1,2} \qquad h^{1,2} \qquad 0 \\
0 \qquad h^{1,1} \qquad 0 \\
0 \qquad 0 \\
1\\
\end{gathered}
\end{equation}
showing that there remains few non-trivial components for consideration in the subsequent string compactification. 
For Calabi-Yau four-folds, the non-trivial Hodge numbers are $\{h^{1,1},h^{1,2},h^{1,3},h^{2,2}\}$. Noting that in this dimension there exists a further constraint on the Hodge numbers \cite{Sethi_1996}, which reads
\begin{align}\label{eq:Constraint}
    -4 h^{1,1}+2 h^{1,2}-4 h^{1,3}+h^{2,2}=44 \; ,
\end{align}
allowing $h^{2,2}$ to be eliminated from the above list.

These Hodge numbers, as well as the Euler number, are of particular interest to physicists; and work characterising and classifying Calabi-Yau's beyond these topological properties has seen insightful early progress \cite{Chandra:2023afu,Gendler:2023ujl}.
For the weighted projective space construction of Calabi-Yau manifolds there are direct formulas for these topological properties from the weights alone \cite{Vafa:1989xc,Batyrev:2020ych}.
Specifically these are
\begin{equation}\label{hodgeeulerformulas}
\begin{split}
    \chi & = \frac{1}{w_{tot}} \sum_{l,r=0}^{w_{tot}-1} \bigg[\prod_{i|lq_i \& rq_i \in \mathbb{Z}} \bigg(1-\frac{1}{q_i}\bigg)\bigg]\;,\\
    Q(u,v) & = \frac{1}{uv} \sum_{l=0}^{w_{tot}} \bigg[ \prod_{\tilde{\theta}_i(l)\in\mathbb{Z}} \frac{(uv)^{q_i}-uv}{1-(uv)^{q_i}} \bigg]_{int} \bigg( v^{size(l)} \bigg(\frac{u}{v}\bigg)^{age(l)}\bigg) \;,
\end{split}
\end{equation}
for weights $w_i$, normalised weights $q_i = w_i/w_{tot}$, and $u, v$ as dummy variables of the  Poincaré polynomial $Q(u,v) := \sum_{p,q} h^{p,q}u^pv^q$. 
For $Q(u,v)$, $\tilde{\theta}_i(l)$ is the canonical representative of $lq_i$ in $(\mathbb{R}/\mathbb{Z})^5$, $age(l) = \sum_{i=0}^4 \tilde{\theta}_i(l)$, and $size(l) = age(l) + age(w_{tot} - l)$.
Note also for $\chi$, where $\forall \, i \ lq_i \ or \ rq_i \notin \mathbb{Z}$ then the product takes value 1.
These components are reintroduced and explained in more detail in §\ref{sec:Approximation}.

These formulas, as can be seen from the equations \eqref{hodgeeulerformulas}, are especially complicated, requiring factorially many integer divisibility checks as the weights in the weight system increase in value, as well as numerous extremely expensive polynomial divisions.
It is with this in mind, that this work is motivated to investigate the efficacy of machine learning methods at approximating these formulas in §\ref{sec:ml}, with the aim of distilling physical insight to form a suitable approximation, as discussed in §\ref{sec:Approximation}; focusing on the more demanding Poincar\'e polynomial formula for Hodge numbers -- from which the Euler number can be computed.

%%%%%%%%%%%%%%%%%%%%%%%%%%%%%%%%%%%%%%%%%%%%%%%%%
\section{Data Analysis}\label{sec:dataanalysis}
In this section, the database of transverse 6-vector weight systems, used to construct Calabi-Yau (CY) four-folds via $\mathbb{P}^5_\textbf{w}$ spaces, is analysed from a general data science perspective.
Databases of weight systems satisfying different combinations of the considered properties for transversality are then generated and discussed, with further data analysis.

\subsection{The Four-folds Dataset}
\begin{figure}[tb]
\includegraphics[scale=0.53]{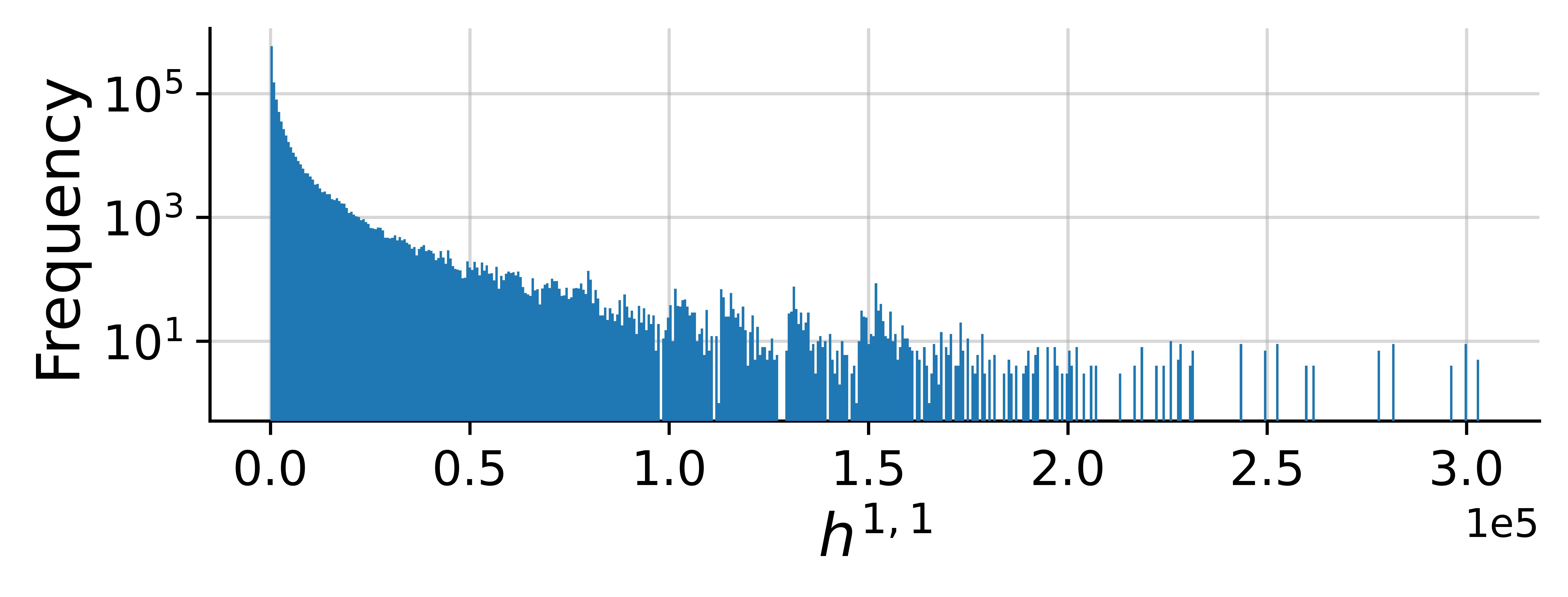}
\includegraphics[scale=0.53]{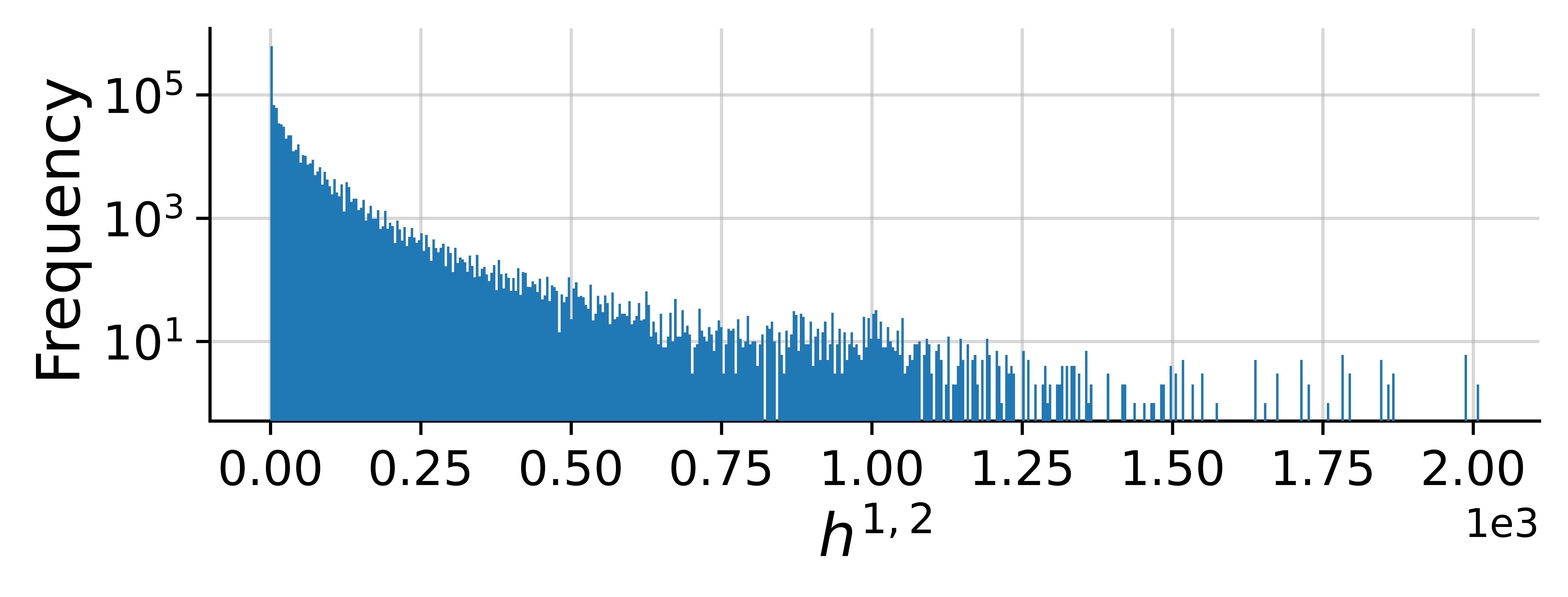}
\includegraphics[scale=0.53]{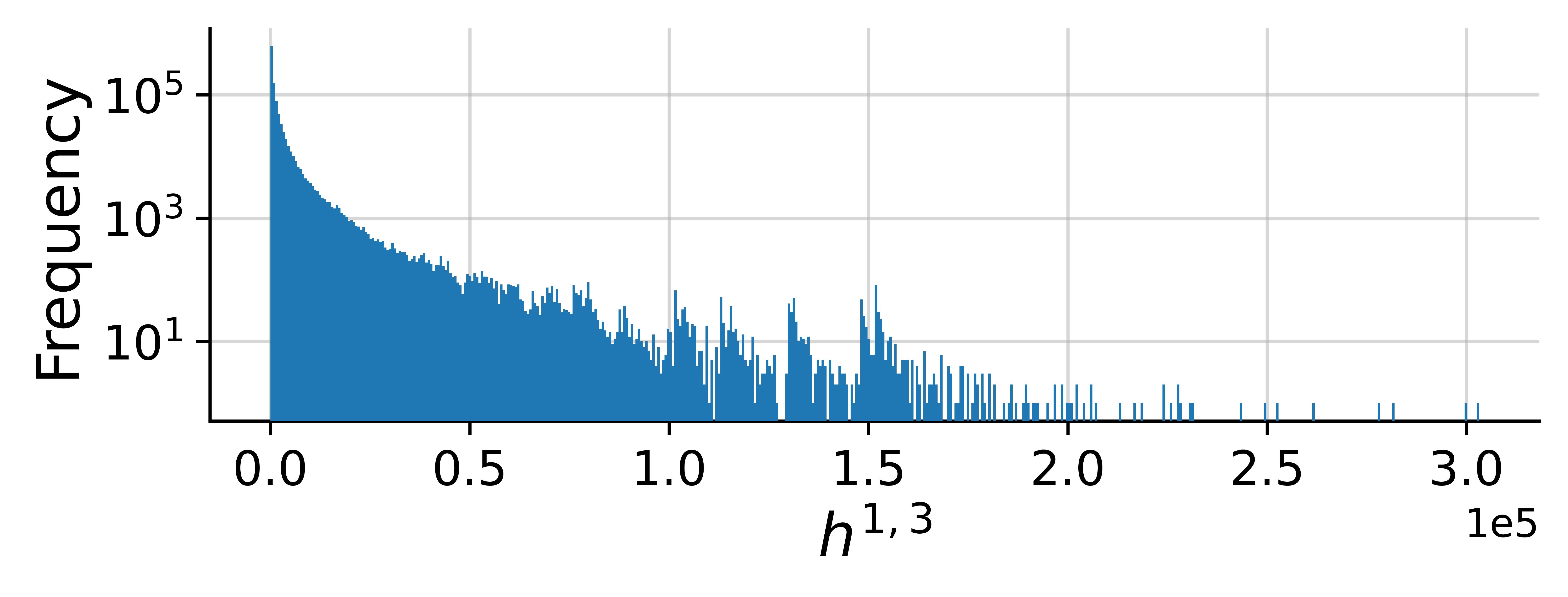}
\includegraphics[scale=0.53]{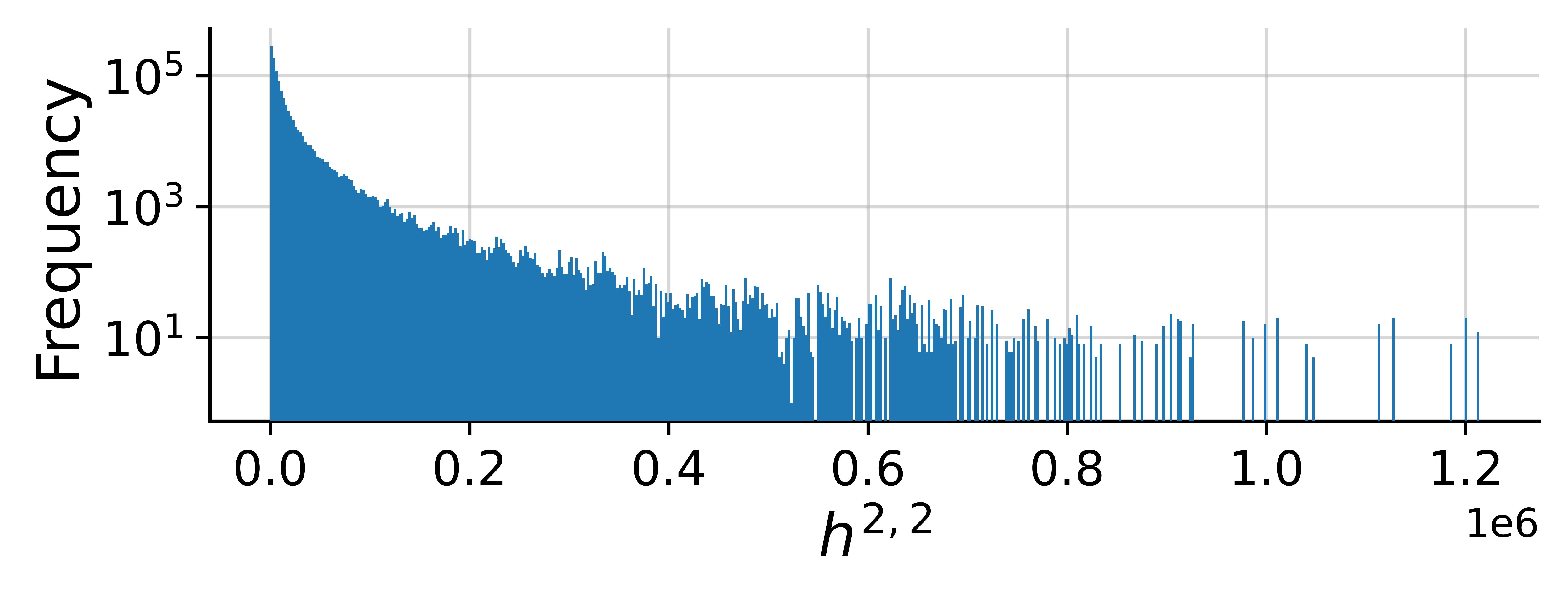}
\centering
\includegraphics[scale=0.53]{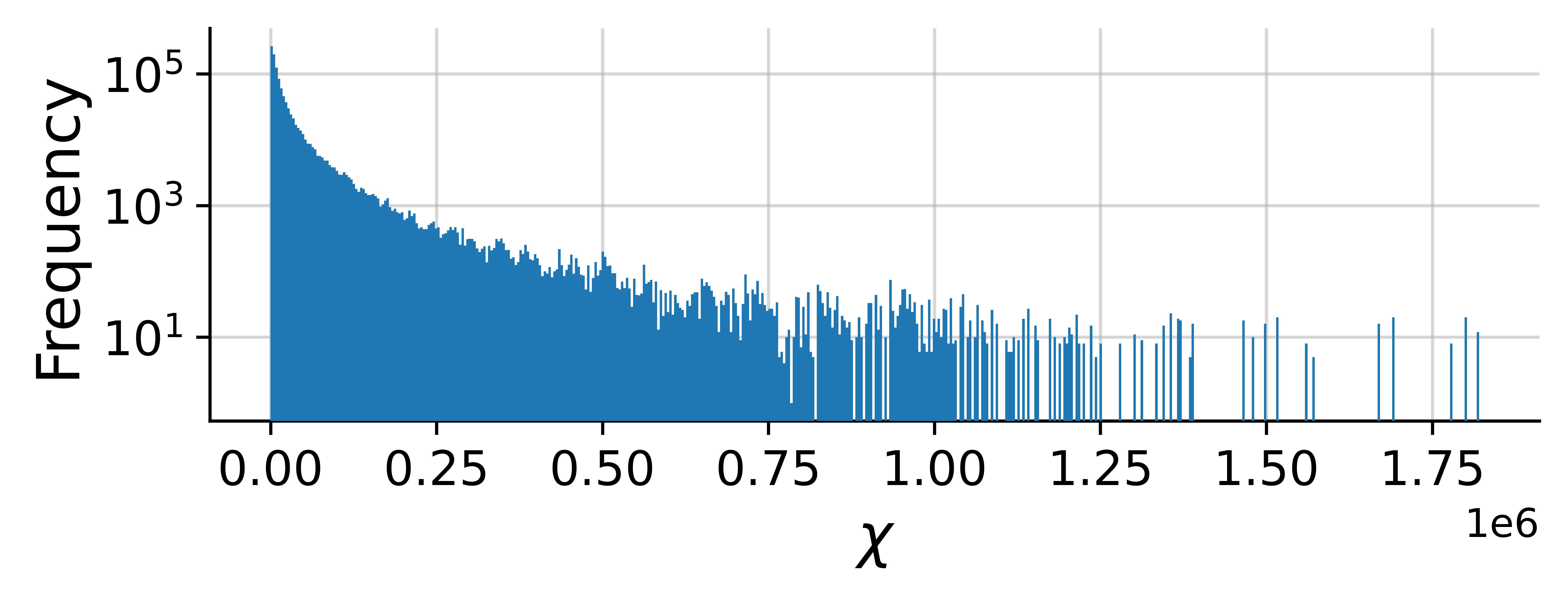}
\caption{These plots illustrate the distribution of each of the invariants for Calabi-Yau four-folds in weighted projective spaces. Each bin contains 500 samples.}
\label{fig:Hist}
\end{figure}

\begin{figure}[tb]
\begin{subfigure}{0.47\textwidth}
\centering    \includegraphics[scale=0.54]{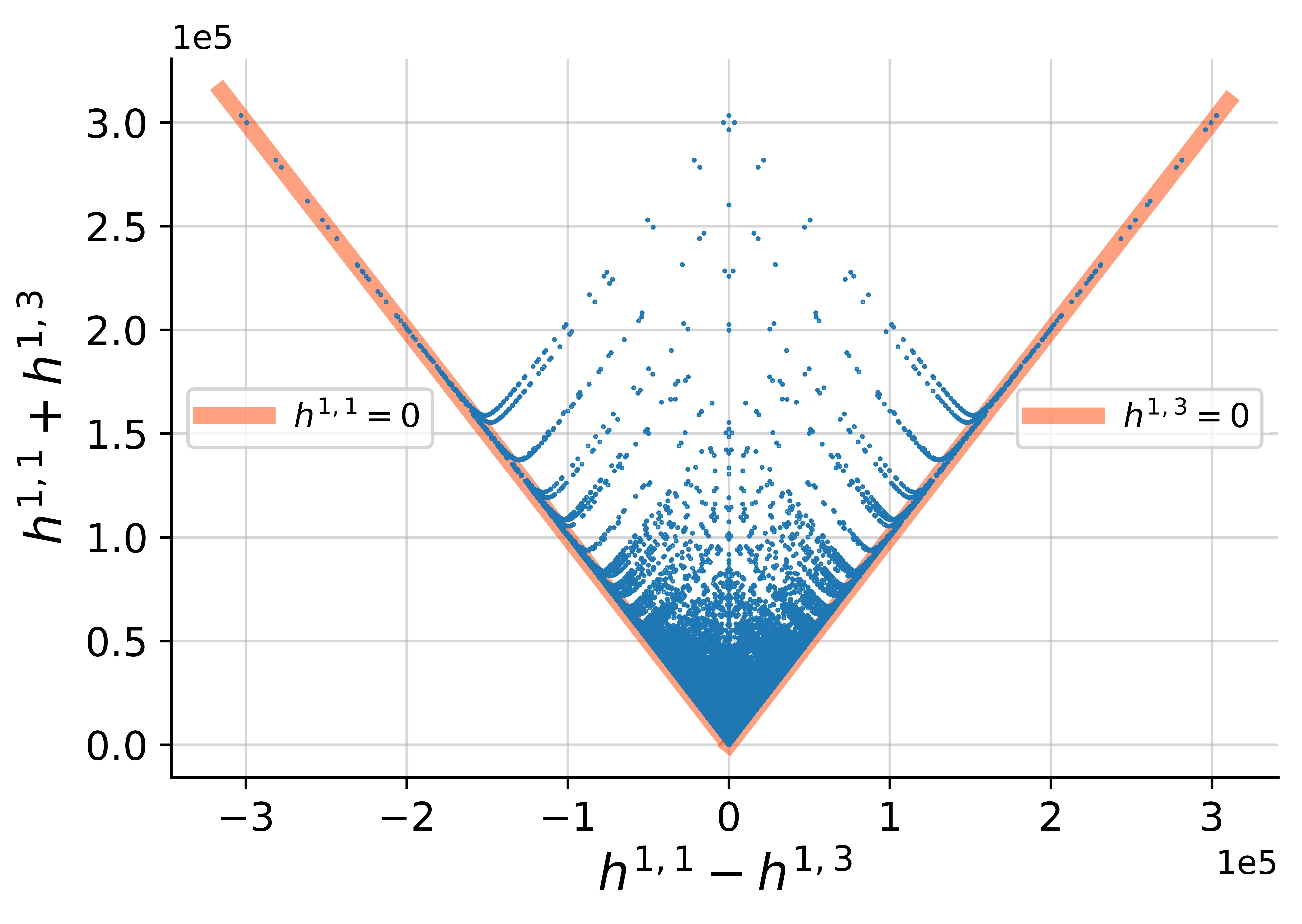}
 \caption{}\label{fig:Symma}
 \end{subfigure}  
 \begin{subfigure}{0.47\textwidth}
\centering   \includegraphics[scale=0.54]{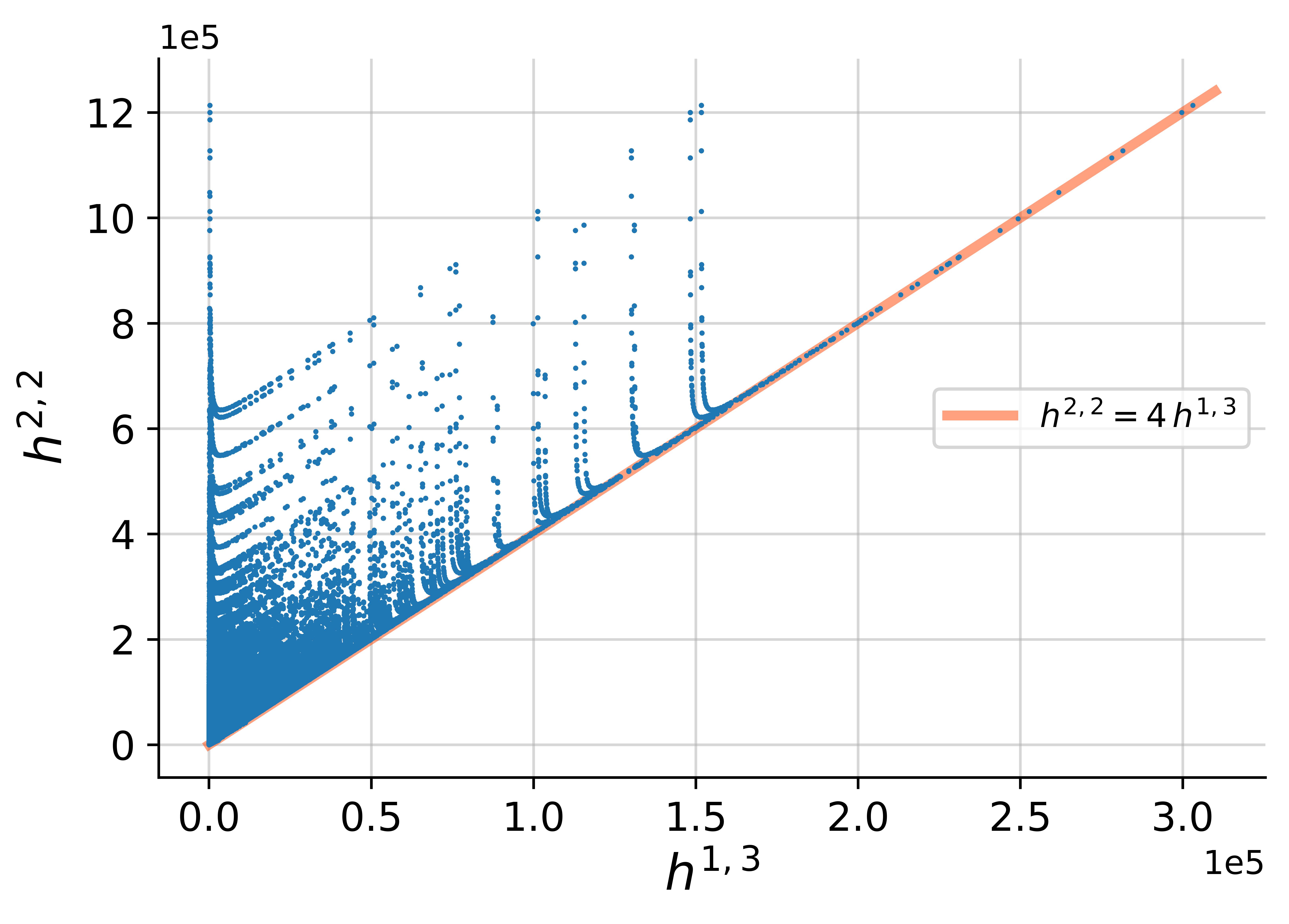}
\caption{}\label{fig:Symmb}
 \end{subfigure} 
\centering
\begin{subfigure}{0.63\textwidth}
\centering 
\includegraphics[scale=0.85]{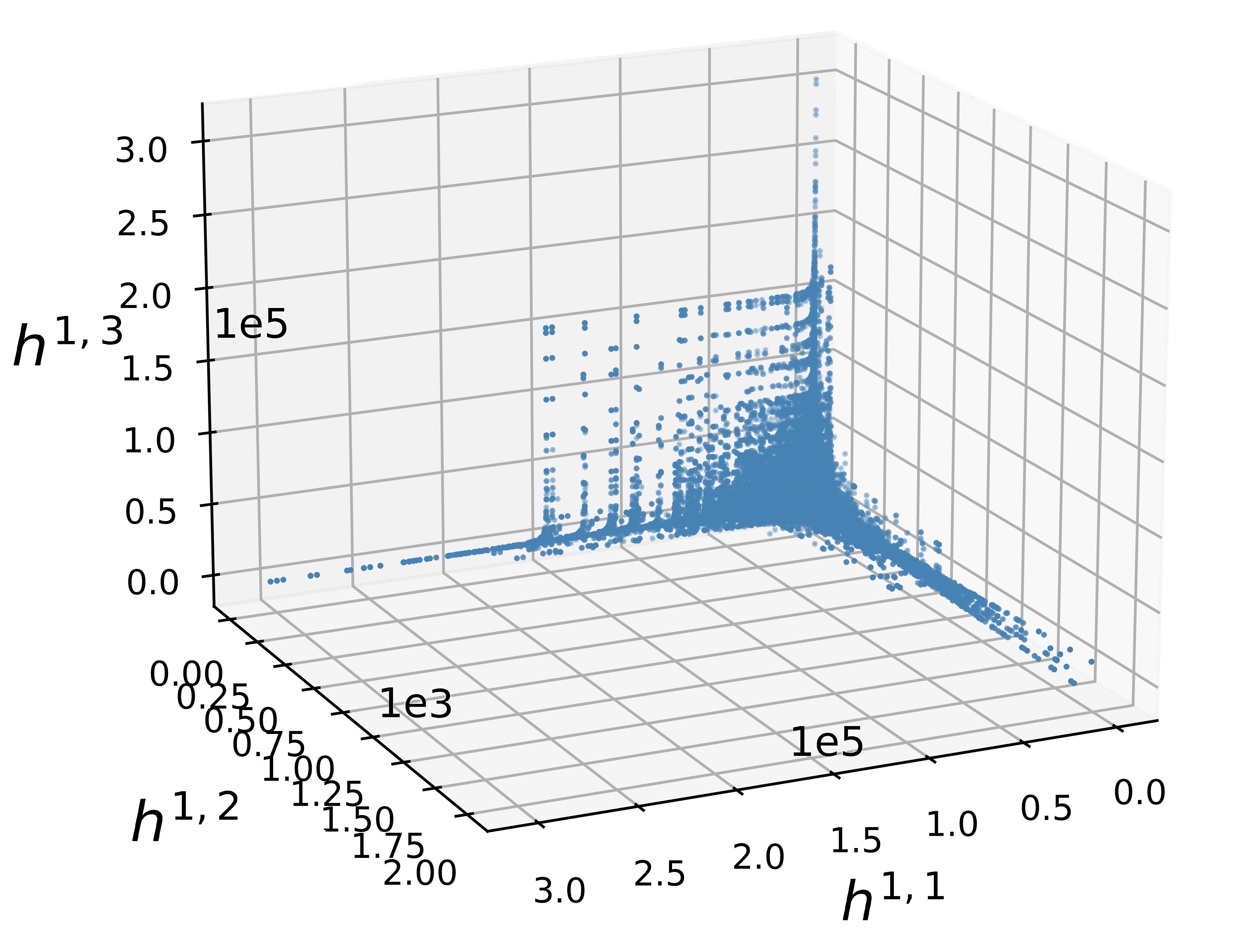}
\caption{}\label{fig:Symmc}
 \end{subfigure} 
    \caption{Scattered plots of the Hodge numbers of Calabi-Yau four-fold's in weighted projective spaces. Plot (a) illustrates that the spaces are mirror symmetric to a high degree. Whilst (b) shows the relation between the two highest Hodge numbers, also compared with the constraint \eqref{eq:Constraint}. Finally, the 3D plot (c) illustrates the relation between the three independent Hodge numbers.}
    \label{fig:Symm}
\end{figure}

Here, the global properties of the primary dataset under investigation are summarised. It was first presented in \cite{Lynker_1999}, where some patterns in the Hodge numbers arrangement were discussed and illustrated by scattered plots; with further preliminary plots available in \cite{He:2015fif}. Our work, on the other hand, is a natural extension of the investigations on the analogous manifolds in three complex dimensions, performed in \cite{Berman_2022}. As such, we focus on the features that are most relevant for machine learning purposes, and we start from the distribution of the invariants, which is shown in Figure \ref{fig:Hist}. 
\graphicspath{ {./Figures/} } 
We observe that, by using the logarithmic scale for the frequency, all histograms display a similar behaviour. The majority of samples is always concentrated around low values, and the ranges span several order of magnitudes. The key features of the distributions in Figure \ref{fig:Hist} can be summarised as
\begin{align}
\left\langle h^{1,1}\right\rangle= 2933.8_{\,1}^{ \, 303148}\,, \quad  \left\langle h^{1,2}\right\rangle=24.1_{\,0}^{2010} \, &, \quad  \left\langle h^{1,3}\right\rangle=2300.3_{\, 1}^{\, 303148}\, ,   \nonumber \\[1em] \left\langle h^{2,2}\right\rangle=20932.0_{\, 82}^{\, 1213644} 
\, , \quad &\langle\chi\rangle=31307.5_{\, -252}^{\, 1820448} \, ,
\end{align}
where we borrow the notation $\mathrm{mean}^{\, \mathrm{max}}_{\, \mathrm{min}}$ from \cite{Gray_2014}. The same range and similar mean values of $h^{1,1}$ and $h^{1,3}$ are a hint of mirror symmetry, which is indeed present in this dataset, as noted in \cite{Lynker_1999}; Figure \ref{fig:Symma} provides an illustration of it. This plot should be compared with the famous three-fold version, where $h^{1,1} + h^{1,2}$ is plotted against the Euler number $\chi$ \cite{CANDELAS1990383}. Quantitatively, the degree of mirror symmetry is around $70 \%$, as reported in \cite{Lynker_1999}. This feature was discovered in generating the set of Calabi-Yau's constructed as hypersurfaces in weighted projective spaces, derived from their embedding within toric varieties, and notably does not apply to the CICY construction \cite{Ashmore_2012,Gray_2014}. 

Figure \ref{fig:Symmb} shows the relation between the two highest Hodge numbers (note that, due to mirror symmetry, this would look almost identical if we were to plot $h^{1,1}$ instead of $h^{1,3}$). The orange line corresponds to $h^{1,1},h^{1,2} << h^{1,3}$, as can be seen from the relation \eqref{eq:Constraint}, and a good amount of data clusters along this line. This feature was noted in \cite{Gray_2014}, where they analysed the less symmetric set of complete intersection Calabi-Yau four-fold Hodge numbers, and found that the data only showed the linear behaviour depicted in the plot (in orange).
The distribution of the non-trivial Hodge numbers $h^{1,\bullet}$ is shown in Figure \ref{fig:Symmc}. By virtue of \eqref{eq:Constraint}, this contains all of the cohomological information of the manifold. As we expect, the  $h^{1,1}$-$h^{1,3}$ plane at $h^{1,2} \approx 0$ displays the mirror-symmetric behaviour shown in Figure \ref{fig:Symma} (with a $45^{\circ}$ rotation). 

\begin{figure}[tb]
\centering
\includegraphics[scale = 0.75]{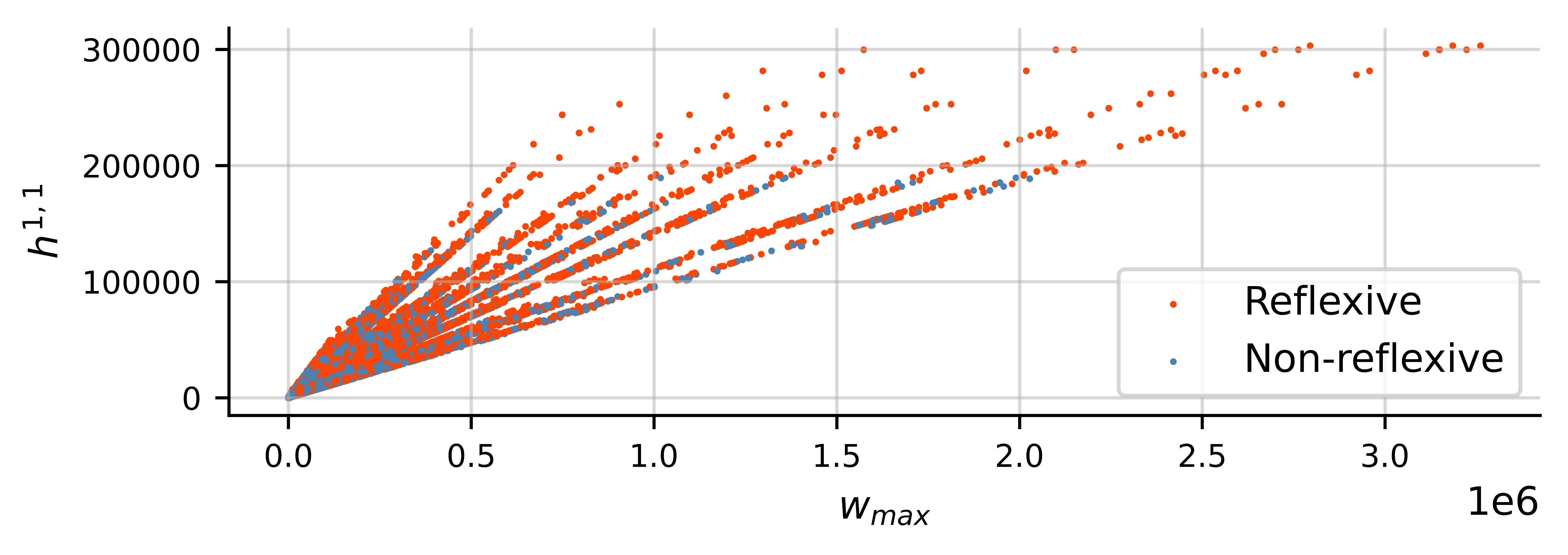}
\caption{This plot of $h^{1,1}$ as a function of the highest weight, $w_{max}$, shows that the linear clustering observed in \cite{Berman_2022} for Calabi-Yau three-folds in weighted projective spaces is also manifest in the four-folds dataset. Both reflexive and non-reflexive systems display the same behaviour, although the regime $h^{1,1} > 200000$ is dominated by reflexive ones. This is confirmed by the principal component analysis shown in Figure \ref{fig:PCA}.}
\label{fig:h11R_vs_nR}
\end{figure}

\begin{figure}[tb]
\centering
\begin{subfigure}{0.47\textwidth}
\centering
\includegraphics[scale=0.5]{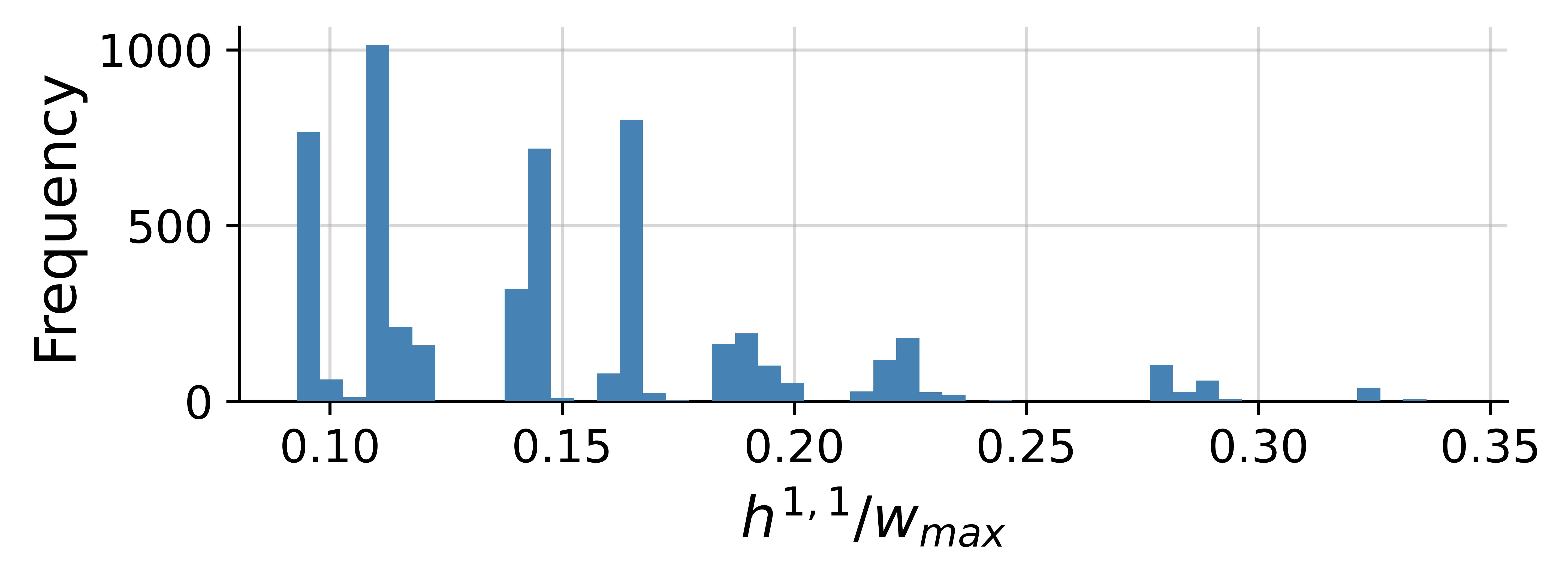}
\caption{}\label{fig:Clusteringa}
\end{subfigure}
\begin{subfigure}{0.47\textwidth}
\centering
\includegraphics[scale=0.5]{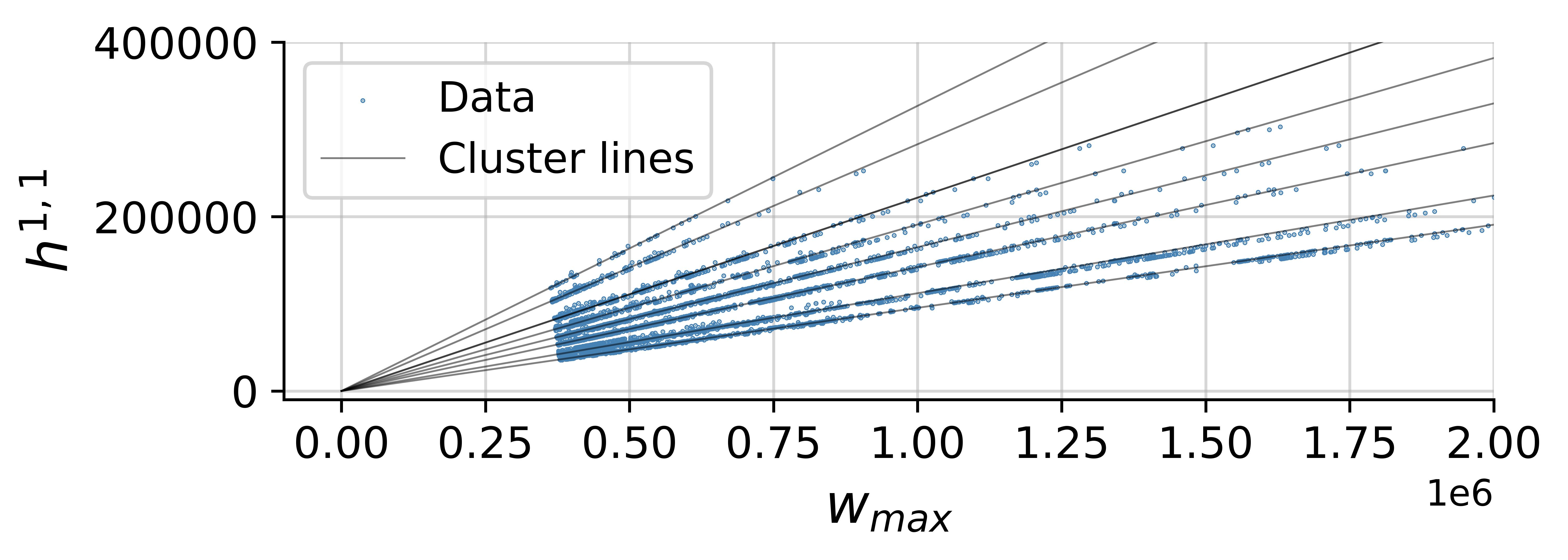}
\caption{}\label{fig:Clusteringb}
\end{subfigure}
\caption{These plots focus on the clustering observed in Figure \ref{fig:h11R_vs_nR} for large weights, and show that the clustering analysis correctly reproduces the multi-linear behaviour. The clusters are shown as peaks in the $h_{1,1} / w_{max}$ histogram in (a) and as lines in the $h_{1,1}$ vs $w_{max}$ plane in (b).}
\label{fig:Clustering}
\end{figure}

Another interesting feature of this dataset is that, analogously to what was observed in \cite{Berman_2022}, an evident linear forking behaviour in the plot of $h^{1,1}$ vs highest weight of the system can be observed. It is shown in Figure \ref{fig:h11R_vs_nR}, where the dataset was also partitioned into reflexive and non-reflexive weight systems.
\graphicspath{ {./Figures/} }
This partitioning is discussed in more detail and put in a broader context in §\ref{sec:Partitioning}. For now, we just note that for highest weights larger than $\sim 5 \times 10^4$, the $h^{1,1}$ values fall neatly into linear clusters. Motivated by the findings presented in \cite{Berman_2022}, we explore this behaviour of the dataset at hand with similar techniques. As it is evident from Figure \ref{fig:h11R_vs_nR}, for large weights, there are eight peaks in the $h^{1,1}/w_{max}$ distribution (where the largest weight is the final weight in the weight system, such that $w_{max}=w_5$). These are linear clusters in the $h^{1,1}$ vs $w_{max}$ plane, as shown in Figure \ref{fig:Clusteringa}. To neatly illustrate the clusters, we only considered systems with largest weights $w_{max}  \gtrapprox 3 \times 10^5$, which can be seen from Figure \ref{fig:Clusteringb}.
The gray lines are the clusters obtained via the K-Means algorithm as used in \cite{Berman_2022}, which is now described.
The statistical confidence of a clustering behaviour can be quantified by the inertia measure. Running the K-Means algorithm on an input dataset with a prespecified number of clusters, the cluster centres/means $\mu_{\mathscr{C}}$ are randomly initialised and the datapoints are allocated to the clusters they are closest to.
In each cluster, the centre is then updated to the mean of the datapoints allocated to it, from which all datapoints are then reallocated to the clusters they are closest to with respect to these new means.
This process is iterated until convergence.
Given a final set of clusters $\mathscr{C}$, with associated means $\mu_{\mathscr{C}}$, across the clustered dataset which here is on inputs $r_i = h^{1,1}/w_{max}$, then the inertia is defined as
\begin{align}
\mathscr{I}=\sum_{\mathscr{C}} \sum_{r_i \in \mathscr{C}}\left(\mu_{\mathscr{C}}-r_i\right)^2 \, .
\label{eq:Inertia}
\end{align}
We are implicitly assuming that any $r_i$ belongs to the cluster who's mean it is closest to. The number of clusters for the problem at hand was found to be $8$, deduced by eye from Figure \ref{fig:Clustering}. Furthermore, two normalised versions of \eqref{eq:Inertia} may also be introduced, which have a nice statistical interpretation:
\begin{align}
    \hat{\mathscr{I}} = \frac{\mathscr{I}}{n_{\mathrm{samples}}}\quad \quad \mathrm{and} \quad \quad \hat{\hat{\mathscr{I}}} = \frac{\hat{\mathscr{I}}}{\mathrm{max}(r_i) - \mathrm{min}(r_i)} \, \, \, .
\end{align}
They are normalised with respect to the number of samples, and with respect to both the number of samples and the range of the samples, respectively. For the clustering analysis of the $h^{1,1}, w_{max}$ data reported in Figure \ref{fig:Clustering}, we find:
\begin{align}
    \mathscr{I} =0.050 \, , \quad \quad \hat{\mathscr{I}} = 9.3 \times 10^{-6} \, , \quad \quad \hat{\hat{\mathscr{I}}} = 3.8 \times 10^{-5} \, .
\end{align}
In words, these value show that, on average, the ratios $h^{1,1}/w_{max}$ that we considered are $0.0038 \%$ of the total range from their nearest cluster (for range shown in Figure \ref{fig:Clusteringa} to be $\approx 0.25$). These results strongly corroborate the liner clustering behaviour observed.

\subsection{Additional Weight Datasets}
\label{sec:Partitioning}
The exact conditions for transversality of a weight system are derived from the use of the transverse polynomials as potentials of Landau-Ginzburg string vacua \cite{Klemm:1992bx}.
These conditions arise from the necessity for the central charge of these theories to be 9 and a subtle application of Bertini’s theorem allowing deformation of polynomials to reduce the singularity structure to exclusively an isolated singularity. %as required st p=0 & dp=0 have no common solutions in transversailty
 
The direct combinatoric interpretation of these conditions in terms of exclusively the weights is unclear, and as demonstrated in \cite{CANDELAS1990383} a first step towards a complete list of necessary and sufficient conditions is provided by the property we dub \textit{intradivisibility}. 
Due to the necessary but insufficient nature of this property, whilst all transverse weight systems will satisfy it, there are many examples of weight systems satisfying it which induce further singularities on their subsequent hypersurfaces preventing them from being Calabi-Yau in nature and hence the weight system transverse.

As well as the intradivisibility property, via works in \cite{Batyrev:1993oya,Kreuzer:1995cd}, there is a further necessary but insufficient property required for a weight system to exhibit a Calabi-Yau hypersurface. 
This property comes from the interpretation of a weight system as a lattice polytope and respectively the weighted projective space as a compact toric variety.
As described in §\ref{sec:background}, the respective lattice polytope must hence have a single interior point (denoted as the IP property) for the subsequent toric variety to exhibit a Calabi-Yau hypersurface.
Additionally, at this dimensionality ($n>4$) the IP polytope no longer needs to be reflexive to exhibit a Calabi-Yau hypersurface, relaxing the necessity for this condition which is essential for the Calabi-Yau construction in lower dimensions.

\begin{figure}[tb]
    \centering
    \begin{subfigure}{0.47\textwidth}
        \centering
        \includegraphics[width=0.98\textwidth]{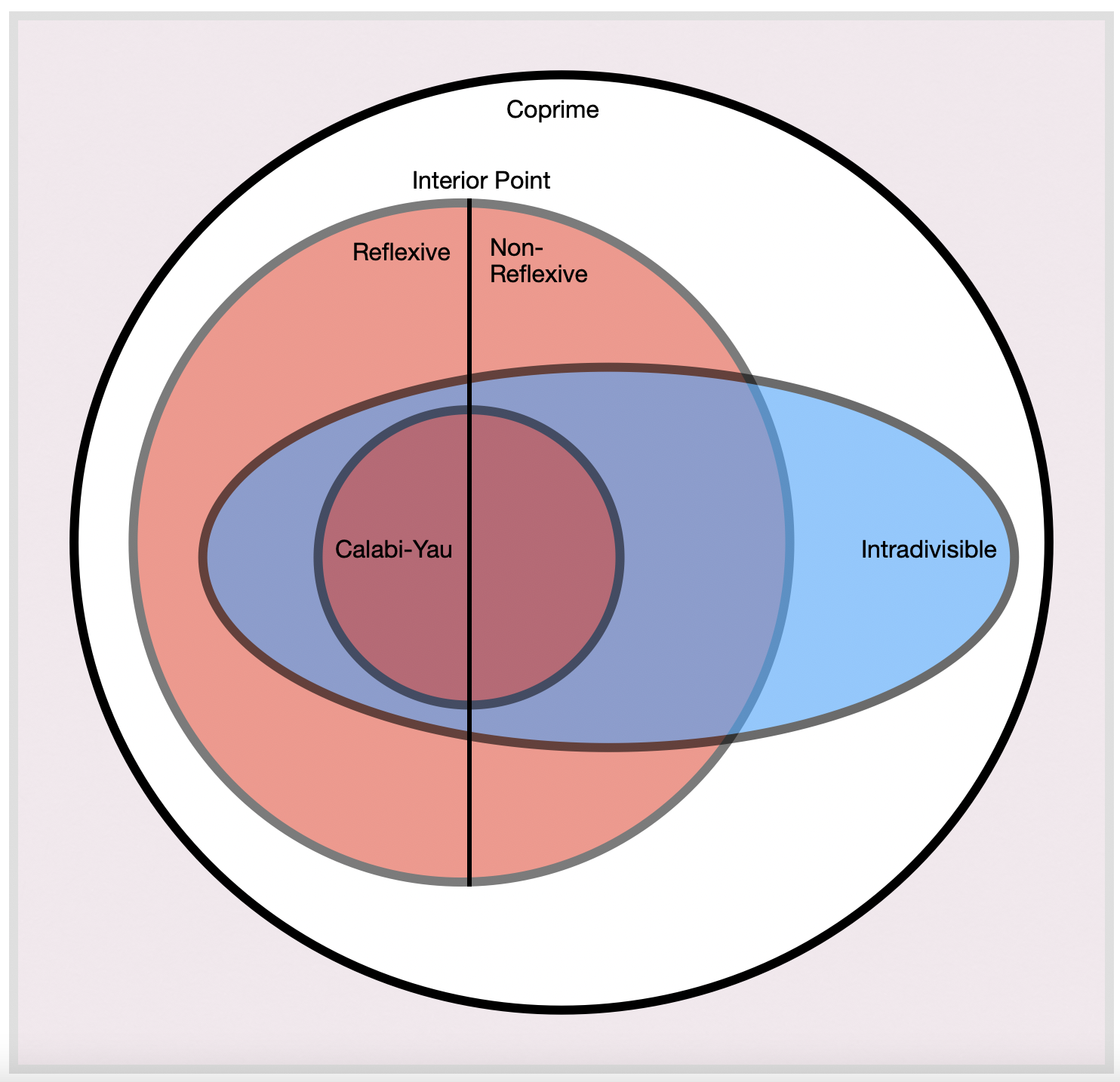}
        \caption{}\label{vennA}
    \end{subfigure} 
    \begin{subfigure}{0.47\textwidth}
        \centering
        \includegraphics[width=0.98\textwidth]{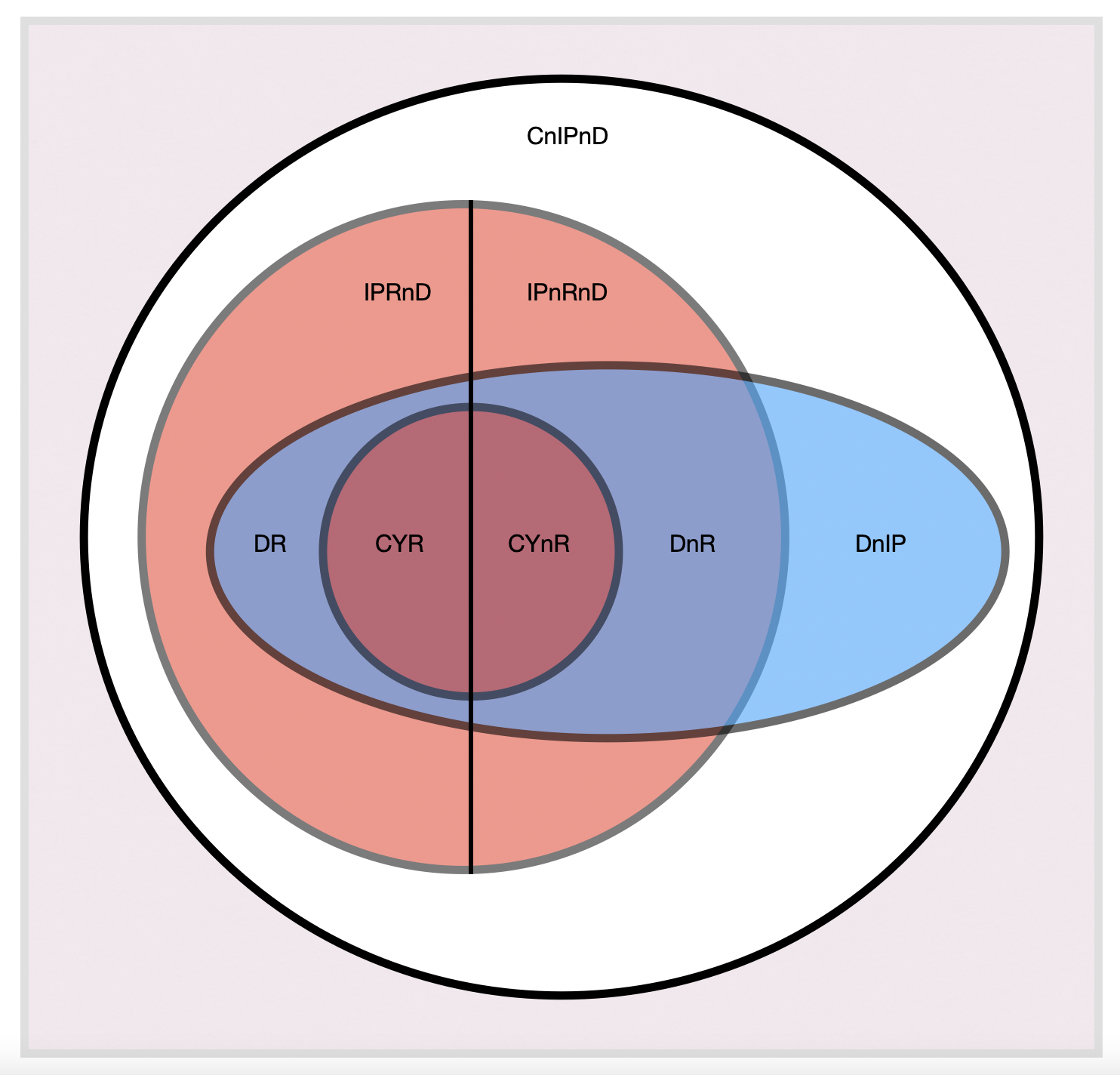}
        \caption{}\label{vennB}
    \end{subfigure} 
    \caption{Venn diagrams displaying (a) the conditional dependencies of the considered 6-weight weight system properties, and (b) the partition of the weight system data into non-overlapping subdatasets.}\label{venndiagrams}
\end{figure}

As demonstrated, for higher-dimensional Calabi-Yau constructions, the relative importance of the previous essential properties becomes less clear; as well as their interrelations.
Therefore, to graphically represent the dependencies of these properties, a Venn diagram is presented in Figure \ref{vennA}.
This in essence classifies the relevant ambient weighted projective spaces, which are defined uniquely by coprime weight systems\footnote{Any common factor can be removed by redefinition of the identification parameter $\lambda$, making the coprime case the natural unique representative of each weighted projective space.}.
Principally, for a 6-vector weight system to be transverse and hence exhibit a Calabi-Yau four-fold hypersurface, it must be \textit{both} IP and intradivisible\footnote{We tested that, as expected, weight systems without the Calabi-Yau property are incompatible with the formula obtained via the Landau-Ginzburg model, described in §\ref{sec:Approximation}. Specifically, the polynomial divisions involved in such an expression are not well defined, i.e. the result contains a reminder.}.

It is therefore interesting to probe this relative importance amongst the necessary conditions using equivalent datasets of weight systems satisfying different combinations of these properties.
Specifically, a dataset of weight systems is constructed for \textit{every} combination of these properties.
The partition of these weight systems is described by starting with coprime weight systems satisfying neither IP or intradivisibility (CnIPnD); then weight systems satisfying either intradivisibility (DnIP), or IP, which can then be non-reflexive (IPnRnD) or reflexive (IPRnD); then weight systems satisfying both intradivisibility and IP, but still not transverse, hence still split into non-reflexive (DnR) and reflexive (DR); and then, finally, the transverse weight systems exhibiting Calabi-Yau hypersurfaces which are again either non-reflexive (CYnR) or reflexive (CYR).
This becomes a full partition of weight systems into subdatasets with respect to these properties, and these subdataset labels are chosen as acronyms to reflect the satisfied properties of coprime $\mapsto$ C, IP $\mapsto$ IP, intradivisible $\mapsto$ D, reflexive $\mapsto$ R, and where relevant the absence of a property is denoted with a `n' before the respective property notation (i.e. non-reflexive $\mapsto$ nR).
This partition is represented on the property interrelation Venn diagram in Figure \ref{vennB}, spanning all unique parts of it.

In generating these datasets, firstly the database of transverse weight systems of \cite{Lynker_1999} was partitioned into reflexive and non-reflexive to produce the CYR and CYnR subdatasets, respectively.
Then the publicly available sample of $1000000$ 5-dimensional 6-vector IP weight systems \cite{Scholler:2018apc} was partitioned into reflexive and non-reflexive as well as intradivisible and non-intradivisible to initiate the respective DR, DnR, IPRnD, IPnRnD subdatasets, explicitly ensuring no overlap with the CYR and CYnR datasets.
New to this work, we generate all intradivisible 6-vector weight systems with sum of weights a maximum value of 400 (using functionality in this paper's \href{https://github.com/Tancredi-Schettini-Gherardini/P5CY4ML}{GitHub}), partitioning off those which are not in the CY property subdatasets and then checking the IP and reflexivity properties (using \texttt{PALP} functionality \cite{Kreuzer_2004}) to supplement the DR, DnR, DnIP subdatasets.
Finally, coprime weight systems were generated stochastically\footnote{In similar vein to \cite{Berman_2022}, an exponential distribution was fitted to the Calabi-Yau weight systems and used to generate trial weight systems, which were checked to be coprime.} and checked for intradivisibility and IP.
These coprime weight systems were then partitioned into IPRnD, IPnRnD, CnIPnD subdatasets, omitting any which were intradivisible to keep the sum of weights maximum value fixed for the DR, DnR, DnIP datasets.
These datasets were then combined with the above, and any repetitions of weight systems removed. 
This substantially increased the subdataset sizes, producing a final partition with class sizes as shown in Table \ref{tab:partitionsizes}.

\begin{table}[tb]
\centering
\addtolength{\leftskip}{-1.75cm}
\addtolength{\rightskip}{-1.75cm}
\begin{tabular}{|c|c|c|c|c|c|c|c|c|}
\hline
Subdataset & CnIPnD & DnIP & IPnRnD & IPRnD  & DnR  & DR    & CYnR   & CYR    \\ \hline
Size       & 408124 & 9614  & 999975 & 988436 & 172462 & 81215 & 847122 & 252933 \\ \hline
$\mathrm{mean^{max}_{min}}$  & $198^{3294}_{1}$ & $39^{198}_{1}$  & $323^{7875}_{1}$ & $59^{858}_{1}$ & $42^{200}_{1}$ & $40^{200}_{1}$ & $3969^{2028138}_{1}$ & $9080^{3260733}_{1}$ \\ \hline
\end{tabular}
\caption{The sizes of the subdatasets of weight systems of 6 weights in each part of the partition; along with the means and ranges across all weight values in each subdataset.}
\label{tab:partitionsizes}
\end{table}

As can be seen in Table \ref{tab:partitionsizes}, the subdatasets are not balanced in size.
In some cases this is particularly natural where the CY subdatasets are exhaustive in their partition between CYnR and CYR, including the entire finite list of possibilities.
Moreover, there are finitely many IP weight systems which can be split amongst the appropriate properties; of which only a sample is publicly available which we supplement with statistical searches. 
Conversely, the intradivisibility property is not expected to enforce finiteness on the dataset of satisfying weight systems. %prove this
Therefore this set has not been generated exhaustively in previous work, and is completed exhaustively here for a sum of weights up to 400\footnote{The sum of weight limit was selected as the limit of computation reached at HPC timeout of 240 core hours.}.
These class sizes are hence well motivated from a viewpoint of exhaustive consideration and analysis, as well as due to computational limitations.
Conversely, there are infinitely many coprime weight systems satisfying neither intradivisibility or IP, which we hence sample stochastically until a suitable order of magnitude matching the other class sizes was achieved.

The difference in subdataset sizes provides concrete stochastic information about the overlap of these weight system properties, and one could then crudely infer probabilities of a generic coprime weight system satisfying each property combination using these dataset sizes.
The later machine learning architectures implemented have generic adaptability to accommodate variable class sizes, as described in §\ref{sec:ml_classify}, and appropriate performance measures are used to avoid bias misinterpretations of learning.

\subsection{Principal Component Analysis}
Linear behaviour in distributions can be analysed through principal component analysis (PCA). 
This unsupervised machine learning technique extracts an orthonormal basis for the dataset in question, with basis vectors ranked according to their degree of contribution towards the variance in the data's distribution.
The basis is computed as eigenvectors of the dataset's covariance matrix, where the symmetric nature of the matrix ensures real eigenvalues that can be ordered decreasingly and then used to rank the basis.
The normalised eigenvalues are named the explained variance, and provide a measure of relative importance of each eigenvector (the larger the explained variance the more important the respective eigenvector).
For a prespecified desired degree of representation, a dataset can be projected onto the first $i$ eigenvectors in the ranked basis such that the sum of the respective first $i$ normalised eigenvalues exceeds the desired proportion of representation.
In this sense, PCA is often used as a dimensionality reduction technique.

In this work, the union of all subdatasets of 6-vector weight systems were analysed with PCA, as one large dataset, to probe the capacity of linear structure being used for simple classification between the subdatasets of the partition.
In this PCA, the explained variances were:
\begin{equation}
    (0.999999498, 0.000885369, 0.000405971, 0.000233533, 0.000010095, 0.000001597) \, , \nonumber
\end{equation}
demonstrating a clear dominance in the first principal component.
Due to the nature of representation of the weight systems, where the entries are sorted in increasing size, it is expected that the latter parts of the vector will dominate the most significant principal components\footnote{Within the method of PCA it is often typical to centre and scale the data components prior to analysis. Centering has no physical effect on the features since the covariance is relative to the mean so is not implemented here, whilst scaling is typically important where each component is a different measure with different units -- not applicable here where there're no units and the relative sizes of the weights are inherently important to the weight system definition.}.
This is the case, as shown by the components of this eigenvector for the first principal component:
\begin{equation}
    (0.000220159, -0.002178435,  0.007423314, -0.006743665,       -0.342499831,  0.939461808) \nonumber \, .
\end{equation}
However the final two components are still of the same magnitude such that the projection is not trivial.
The dominance of the first principal component motivates a 1-dimensional projection of the data, using the above eigenvector.
The $\mathrm{mean}^{\, \mathrm{max}}_{\, \mathrm{min}}$ values for the 1-dimensional projections of each subdataset were:
\begin{equation}\label{eq:PCA_mmm}
\begin{split}
    \langle \text{CnIPnD} \rangle & = 356_{17}^{3041}\,, \qquad\quad\ \langle \text{DnIP} \rangle = 68_{1}^{167}\,, \\
    \langle \text{IPnRnD} \rangle & = 592_{16}^{6313}\,, \qquad\ \ \langle \text{IPRnD} \rangle = 122_{11}^{614}\,, \\ 
    \langle \text{DnR} \rangle & = 78_{2}^{171}\,, \qquad\qquad\ \langle \text{DR} \rangle = 75_{2}^{169}\,, \\
    \langle \text{CYnR} \rangle & = 8318_{2}^{1439056}\,, \qquad \langle \text{CYR} \rangle = 19116_{1}^{2313640}\,,
\end{split}
\end{equation}
given to the nearest integer.
They display similar lower bounds throughout, whilst higher mean and maximum values for non-reflexive subdatasets, and substantially larger ranges for the CY data.

\begin{figure}[tb]
    \centering
    \begin{subfigure}{0.47\textwidth}
        \centering
        \includegraphics[width=0.98\textwidth]{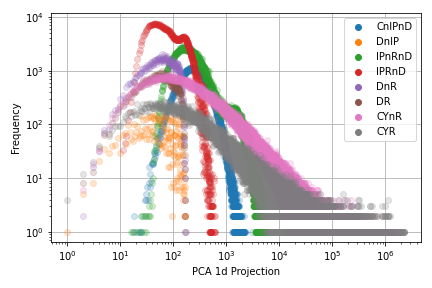}
        \caption{}\label{fig:PCA_all}
    \end{subfigure} 
    \begin{subfigure}{0.47\textwidth}
        \centering
        \includegraphics[width=0.98\textwidth]{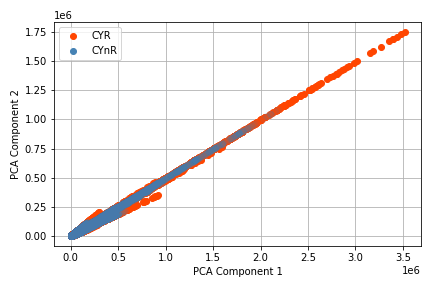}
        \caption{}\label{fig:PCA_CY}
    \end{subfigure} 
    \caption{PCA of the classification partition of weight system data; explained variance demonstrated a single dominant principal component. In (a) the frequency distribution for the 1-dimensional projections of the weight systems are shown for each part of the partition (on a log-log scale). Equivalently (b) shows a 2-dimensional projection of the Calabi-Yau data, corroborating the forking behaviour observed.}\label{fig:PCA}
\end{figure}

To explore the distributions of these projections, their nearest integer values of the 1-dimensional projections for each subdataset in the partition were plotted according to their frequencies of occurrence in the histogram of Figure \ref{fig:PCA_all}.

Plotted according to a log-log scale, the distributions show a surprising approximate continuity of the lines. %interpret
The projection distributions all experience significant overlap in the values they can take, and the overlap of the distribution lines shows that each subdataset experiences regions of the data space where the constituent vectors are distributed similarly to another subdataset (since frequencies are the same in that range of the projections), making classification difficult in each case of comparison.

Interestingly, the distributions of the CY subdataset projection interpolates between the IP and D datasets in the region of highest frequency, respecting this overlap behaviour in the full weight system generation where CY weight systems must be both D and IP.
Alternatively, in each case a property's subdataset is split into R and nR the subsequent subdatasets exhibit distributions of a similar shape.
The non-reflexive cases then have a higher density of high frequencies matching their usually more populous subdatasets distributed over smaller ranges as demonstrated in \eqref{eq:PCA_mmm}.

The similarity between the R and nR subdatasets' PCA projections indicates classification architectures will likely find identification of this property harder.
Whilst the higher skewed values of PCA projection, as well as the far larger maximum values, for the CY datasets will perhaps be used by the architectures to aid learning.

In addition, PCA is performed independently for the dataset of transverse CY weight systems (i.e. union of the CYR \& CYnR subdatasets); exhibiting comparable explained variances and dominant normalised eigenvector.
The 2-dimensional projection of this data is presented in Figure \ref{fig:PCA_CY}, and shows a similar forking structure to Figure \ref{fig:h11R_vs_nR}, equivalently seen for CY three-folds in \cite{Berman_2022}.
Furthermore, as seen in the plots against $h^{1,1}$, the reflexive weight systems dominate the tails of the forks.
All these comparisons corroborate the suggested intimately linear relationship between the weights and $h^{1,1}$, priming the data for machine learning application.

%%%%%%%%%%%%%%%%%%%%%%%%%%%%%%%%%%%%%%%%%%%%%%%%%
\section{Machine Learning}\label{sec:ml}
In this section we present the results of various investigations performed through supervised machine learning (ML). Neural networks (NNs) are employed to predict the Hodge numbers of Calabi-Yau four-folds and to identify weight systems with specific properties. These two applications are different in nature, and for this reason, despite using the same NN architecture, some of the meta-data choices differ. 

NNs are high-dimensional non-linear function fitters, they are built from constituent neurons which receive a vector input, act linearly on that vector to produce a number, then act non-linearly on that number with an activation function: $\underline{x} \mapsto act(\underline{\mathfrak{w}}\cdot \underline{x} + \mathfrak{b})$, for NN weights $\mathfrak{w}$ (not to be confused with weight system weights $w_i$), bias $\mathfrak{b}$, and activation $act(\cdot)$.
The neurons are organised into layers, such that the output numbers of each neuron in a layer are concatenated into a vector to pass to all the neurons in the next layer.
Over training the \textit{optimiser} compares output predictions of the NN function to true values of training data through a \textit{loss}, updating the $(\mathfrak{w},\mathfrak{b})$ parameters to optimise the fitting.
After training is complete the trained NN is used to predict output values on independent test data, from which performance measures can then be calculated \cite{Ruehle:2020jrk}.

For the prediction of cohomological data, which has a very wide range of possible values, a NN regressor was used. Since the input data are small (just six integers), a simple architecture with few layers was enough for this problem. 
Specifically, we used the built-in multi-layer perception regressor from \texttt{scikit-learn}, with the following features: $(16, 32, 16)$ layer structure, ReLU activation, $MSE$ loss, and Adam optimiser. We chose a training-test split of $80:20$, and performed a 5-fold cross-validation for each investigation. The batch size was set to $200$, and we imposed an upper bound of $250$ epochs (the network could stop before that if it reached convergence). Regarding the performance measures, we focused on the following three:
\begin{alignat}{3}
& MSE &&=\frac{1}{n} \sum\left(y_{\text {pred }}-y_{\text {true }}\right)^2 \quad &&\in[\mathbf{\mathbf{0}}, \infty)\;, \nonumber \\
& MAPE &&=\frac{1}{n} \sum\left|\frac{y_{\text {pred }}-y_{\text {true }}}{y_{\text {true }}}\right| \quad &&\in[\mathbf{\mathbf{0}}, \infty)\;, \\
& R^2 &&=1-\frac{\sum\left(y_{\text {true }}-y_{\text {pred }}\right)^2}{\sum\left(y_{\text {true }}-y_{\text {truemean }}\right)^2} &&\in(-\infty, \mathbf{\mathbf{1}}]\;, \nonumber
\end{alignat}
for outputs $y$, where the bold numbers indicate the optimal values, i.e. those corresponding to perfect prediction.

Conversely, for the identification of weight system properties a NN classifier was employed.
This was built and implemented with the same architecture as the regressor (using \texttt{tensorflow} \cite{tensorflow2015-whitepaper}), however changing the loss and performance measures to match the classification problem style.
The loss function was categorical cross-entropy, and performance measures were functions of the confusion matrix.
A confusion matrix, $CM_{ij}$, counts the number of test data inputs in class $i$ that the trained NN classifies into class $j$, this can then be normalised, and the performance measures defined:
\small
\begin{alignat}{3}
    & Accuracy &&= \quad \sum_i CM_{ii} \quad \in [0,\textbf{1}]\;,\\
    & MCC &&= \frac{\sum_{ijk}\big(CM_{ii}CM_{jk}-CM_{ij}CM_{ki}\big)}{\sqrt{\big(\sum_i\big( \sum_j CM_{ij} \big)\big( \sum_{k\neq i,l} CM_{kl}\big)\big)\big(\sum_i\big( \sum_j CM_{ji} \big)\big( \sum_{k\neq i,l} CM_{lk}\big)\big)}} \in [-1,\textbf{1}]\;,\nonumber
\end{alignat}
\normalsize
again where bold values indicate optimal values for perfect learning.
Note that where accuracy is very interpretable as the proportion of correctly classified inputs, the Matthew's correlation coefficient (MCC) is in general a more representative measure as it accounts for off-diagonal terms and hence generalised Type I and II errors.

\subsection{Regressing Hodge Numbers}
Following the promising performances presented in \cite{Berman_2022}, we employ a supervised ML technique on the Calabi-Yau weight system dataset under investigation. While for the three-folds case both $h^{1,1}$ and $h^{1,2}$ could be learned to high levels of precision, we find that the same is not true for four-folds. This does not come as a surprise, since the underlying geometric structure becomes richer and more complicated by going up in complex dimensions. In fact, $h^{1,1}$ is still learnt with very high precision and accuracy, while the architecture described above proves less adequate for $h^{1,2}$ and $h^{1,3}$. This is partially shown in Table \ref{tab:1}, where we also observe a trend that is common to all of our findings.  
\begin{table}[tb]
\begin{tabular}
{|c|c|c|c|c|c|c|c|}
\hline
 Data & \multicolumn{3}{|c|}{$h^{1,1}$} & \multicolumn{3}{|c|}{$h^{1,3}$} \\
\cline { 2 - 7 } Investigated
 & $1^{\mathrm{st}}$ half & $2^{\mathrm{nd}}$ half & Whole & $1^{\mathrm{st}}$ half & $2^{\mathrm{nd}}$ half & Whole \\
 \hline
 \hline \multirow{2}{*}{$R^2$} &  0.9261 & 0.9114 & 0.9101 & 0.9378 & 0.279 & 0.063 \\
& $\pm$ 0.0018 & $\pm$ 0.0024 & $\pm$ 0.0049 & $\pm$ 0.0086 & $\pm$ 0.091 & $\pm$ 0.076  \\
\hline \multirow{2}{*}{\text { MAPE }} & 0.1578 & 0.2498 & 0.409 & 1.68 & 4.50 & 3.03 \\
& $\pm$ 0.0056 & $\pm$ 0.0015 & $\pm$ 0.045 & $\pm$ 0.35 & $\pm$ 0.72 & $\pm$ 0.63 \\
\hline \multirow{2}{*}{\text { MSE }} & 2072 & 14519431 & 8066404 & 1188821 & 61416783 & 48040860 \\
& $\pm$ 55 & $\pm$ 496316 & $\pm$ 550014 & $\pm$ 90490 & $\pm$ 6406298 & $\pm$ 3496383  \\
\hline
\end{tabular}
\caption{This table shows the performances of the fully connected neural network on $h^{1,1}$ and $h^{1,3}$ for the full dataset, the lower half (which contains smaller weights) and the upper half (containing larger weights).}
\label{tab:1}
\end{table}
We note that the small-weights regime is essentially different from the large-weights regime in terms of ML performance. The neural networks yield consistently better results when restricted to the first half of the dataset, compared to the second half. This suggests that there are some features, associated to the large-weights behaviours, which are harder to learn with our architecture\footnote{One might think that this is motivated by the fact that the second half of the dataset contains a wider range of cohomological numbers, since it has a wider range of weights. However, this is not the case, as shown in Table \ref{tab:2}.}. Moreover, we observe another drop in accuracy when investigating the whole dataset, showing that the NN struggles to deal with these two regimes at once. For reference, we report the properties of the two halves in Table \ref{tab:2}.
\begin{table}[tb]
\centering
    \begin{tabular}{|c|c|c|c|}
    \hline
    \multicolumn{2}{|c|}{Subset} & $1^{\mathrm{st}}$ half & $2^{\mathrm{nd}}$ half \\
    \hline \hline
    \multirow{3}{*}{$h^{1,1}$} & Min & 1 & 212  \\
    \cline { 2 - 4 }
    & Max & 1173 & 303148  \\ \cline { 2 - 4 }
    & Mean & 204.1 & 5663.5  \\
    \hline
    \multirow{3}{*}{$h^{1,2}$} & Min & 0 & 0 \\
    \cline { 2 - 4 }
    & Max & 1989 & 2010  \\ \cline { 2 - 4 }
    & Mean & 21.9 & 26.4  \\
    \hline
    \multirow{3}{*}{$h^{1,3}$} & Min & 1 & 1  \\
    \cline { 2 - 4 }
    & Max & 303148 & 227486  \\ \cline { 2 - 4 }
    & Mean & 1475.9 & 3124.7  \\
    \hline
    \end{tabular}
    \quad \quad \quad
    \begin{tabular}{|c|c|c|c|}
    \hline
    \multicolumn{2}{|c|}{Subset} & $1^{\mathrm{st}}$ half & $2^{\mathrm{nd}}$ half \\
    \hline \hline
    \multirow{3}{*}{$h^{2,2}$} & Min & 82 & 1062  \\
    \cline { 2 - 4 }
    & Max & 1213644 & 1213644  \\ \cline { 2 - 4 }
    & Mean & 6720.0 & 35143.9  \\
    \hline
    \multirow{3}{*}{$\chi$} & Min & -252 & 720 \\
    \cline { 2 - 4 }
    & Max & 1820448 & 1820448  \\ \cline { 2 - 4 }
    & Mean & 9996.2 & 52618.8  \\
    \hline
    \multirow{3}{*}{$w_{tot}$} & Min & 6 & 4480 \\
    \cline { 2 - 4 }
    & Max & 4480 & 6521466  \\ \cline { 2 - 4 }
    & Mean & 1561.6 & 60170.4  \\
    \hline
    \end{tabular}
    \caption{These tables show the ranges of the topological quantities under investigation for the two halves of the dataset, and their mean value. The dataset is ordered according to the sum of weights (see bottom of the right table), and not according to any of the cohomological properties. Hence, we see that the range of the invariants does not split into two disjoint sets among the two halves.}
    \label{tab:2}
\end{table}
In order to probe the performance of ML on the full problem, i.e. determining the complete Hodge diamond, we also focused on learning $h^{2,2}$; both on its own and together with the two Hodge numbers above. As shown in §\ref{sec:background}, such a triple is enough to contain all the cohomological information. The results of these investigations are shown in Table \ref{tab:3}.
\begin{table}[tb]
\centering
\addtolength{\leftskip}{-1.75cm}
\addtolength{\rightskip}{-1.75cm}
\begin{tabular}
{|c|c|c|c|c|c|c|c|}
\hline
 Data & \multicolumn{3}{|c|}{$h^{2,2}$} & \multicolumn{3}{|c|}{$(h^{1,1}, h^{1,3}, h^{2,2})$} \\
\cline { 2 - 7 } Investigated
 & $1^{\mathrm{st}}$ half & $2^{\mathrm{nd}}$ half & Whole & $1^{\mathrm{st}}$ half & $2^{\mathrm{nd}}$ half & Whole \\
 \hline
 \hline \multirow{2}{*}{$R^2$} &  0.944 & 0.714 & 0.6228 & 0.670 & 0.529 & 0.528 \\
& $\pm$ 0.015 & $\pm$ 0.022 & $\pm$ 0.0082 & $\pm$ 0.093 & $\pm$ 0.016 & $\pm$  0.015  \\
\hline \multirow{2}{*}{\text { MAPE }} & 0.497 & 0.60 & 0.67 & 1.9 & 3.3 & 2.0 \\
& $\pm$ 0.060 & $\pm$ 0.09 & $\pm$ 0.04 & $\pm$ 0.4 & $\pm$ 0.4 & $\pm$ 0.2 \\
\hline \multirow{2}{*}{\text { MSE }} & 19287530 & 1295749477 & 1006910414 & 37962157 & 578685604 & 348268647 \\
& $\pm$ 5619861 & $\pm$ 109548762 & $\pm$ 36669624 & $\pm$ 15616868 & $\pm$ 25936516 & $\pm$ 10385842  \\
\hline
\end{tabular}
\caption{This table shows the performances of the fully connected neural network on $h^{2,2}$ and on the triple ($h^{1,1}$,$h^{1,3}$,$h^{2,2}$), which specifies all the cohomological information. Again, we report results associated to the full dataset, to the lower half only (which contains smaller weights) and to the upper half only (containing larger weights).}
\label{tab:3}
\end{table}
We again see that the accuracy drops from left to right, according to the chosen subset.
Finally, for completeness, we also present our results on $h^{1,2}$ and $\chi$ in Table \ref{tab:4}. Since both of them can be zero, the MAPE measure does not apply to these cases, and therefore we omit it.

\begin{table}[tb]
\centering
\addtolength{\leftskip}{-1.75cm}
\addtolength{\rightskip}{-1.75cm}
\begin{tabular}
{|c|c|c|c|c|c|c|c|}
\hline
 Data & \multicolumn{3}{|c|}{$\chi$} & \multicolumn{3}{|c|}{$h^{1,2}$} \\
\cline { 2 - 7 } Investigated
 & $1^{\mathrm{st}}$ half & $2^{\mathrm{nd}}$ half & Whole & $1^{\mathrm{st}}$ half & $2^{\mathrm{nd}}$ half & Whole \\
\hline \multirow{2}{*}{$R^2$} & 0.9400 & 0.653 & 0.616 & 0.0715 & 0.0554 & 0.0436 \\
& $\pm$ 0.0033 & $\pm$ 0.015 & $\pm$ 0.010 & $\pm$ 0.0069 & $\pm$ 0.0064 & $\pm$ 0.0038  \\
\hline \multirow{2}{*}{\text { MSE }} & 46999083 & 3551553463 & 2309837118 & 2520  & 5100 & 3834 \\
& $\pm$ 2865647 & $\pm$ 143894491 & $\pm$ 77130429 & $\pm$ 80 & $\pm$ 103 & $\pm$ 120  \\
\hline
\end{tabular}
\caption{This table shows the performances of the fully connected neural network on $\chi$ and on the triple $h^{1,2}$. Both invariants can be zero, so we omit the MAPE measure, which is not well-defined.}
\label{tab:4}
\end{table}
%\begin{align}
%\begin{array}{|c|c|c|c|c|c|c|}
%\hline \multirow{2}{*}{\text { Measure }} & \multicolumn{6}{|c|}{\text { Property }} \\
%\cline { 2 - 7 } & \multicolumn{3}{|c|}{\chi} & \multicolumn{3}{|c|}{(\chi, h^{1,1}, h^{1,3})} \\
%\hline  \,\, \text{Dataset} & 1^{\mathrm{st}} \, \text{half} & 2^{\mathrm{nd}} \, \text{half} & \text{Whole} & 1^{\mathrm{st}} \, \text{half} & 2^{\mathrm{nd}} \, \text{half} & \text{Whole} \\
%\hline \multirow{2}{*}{$R^2$} & 0.9400 & 0.653 & 0.616 & 0.8588 & 0.633 & 0.566 \\
%& \pm 0.0033 & \pm 0.015 & \pm 0.010 & \pm 0.0092 & \pm 0.035 & \pm 0.029  \\
%\hline \multirow{2}{*}{\text { MAPE }} & 205497298382739 & 0.3914 & 349930692100927 & 110744511823927 & 1.87 & 91880309940604 \\
%& \pm 95825196804929 & \pm 0.0075 & \pm 165817661610320 & \pm 33445314290361 & \pm 0.33 & \pm 82180247528375 \\
%\hline \multirow{2}{*}{\text { MSE }} & 46999083 & 3551553463 & 2309837118 & 15048200 & 1016694483 & 673997457 \\
%& \pm 2865647 & \pm 143894491 & \pm 77130429 & \pm 2675886 & \pm 72984373 & \pm 50373284  \\
%\hline
%\end{array}
%\end{align}
The fact that higher cohomologies in Calabi-Yau four-folds are harder to learn with neural networks has already appeared in the literature, in \cite{He:2020lbz}. Although they analysed a different construction of four-folds, i.e. complete intersection Calabi-Yau's (CICY), their results also indicate that $h^{1,1}$ is the only Hodge number that can be successfully learnt to high levels of precision with fully connected networks. 
Convolutional neural network variants have exhibited the highest accuracies on the CICY matrix inputs \cite{Erbin:2020srm}, however due to the permutation symmetry of the configuration matrices, as well as the weight system vectors for the construction considered here, the benefits of the convolutional architecture's focus on local properties is lost, we therefore stick to the more general dense feed-forward architectures.

\subsubsection{NN Gradient Saliency}\label{sec:h11saliency}
Some first steps towards interpretability of these NN results starts with gradient saliency analysis.
The trained NNs are (highly-nonlinear) functions from inputs to outputs, differentiating these functions with respect to each of the inputs can give some indication of the dependency of the output classification on each part of the weight system.

\begin{figure}[t]
    \centering
\includegraphics[width=0.55\textwidth]{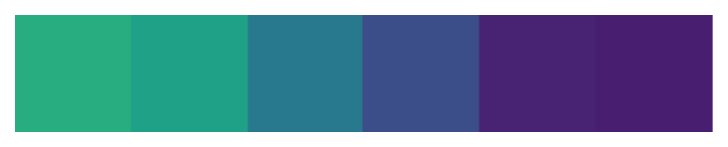}
    \caption{NN gradient saliency scores for the $h^{1,1}$ supervised learning on input weight systems. The lighter colours indicate a larger normalised absolute gradient for that weight in the input 6-vector weight systems, where the saliency scores are averaged over the full test sets of each investigation and each of the 100 repetitions of the investigations.}
    \label{fig:saliency_h11}
\end{figure}

In the saliency analysis performed here, each NN is differentiated with respect to each of the inputs, and the differential evaluated at each of the test data inputs.
The absolute values of these gradient components are then averaged over the test dataset, as well as averaged over the run repetitions -- here repeating the investigation with randomised 80:20 train:test splits for 100 independent NNs of the same architecture.
Since function scales can vary through the NN layers, the relative saliency values are the features of interest; they are represented, for the NNs predicting $h^{1,1}$, in Figure \ref{fig:saliency_h11}.
The 6 weights of the input weight systems are represented by 6 boxes, where lighter colours indicate higher saliency values and larger relative importance.
These results show that the NNs focus on the weights in each system according to their size. They prioritise the information encoded in the lower weights, while the largest weights seem not to play an as important role. This implies that the networks are not exploiting the clustering behaviour shown in Figure \ref{fig:Clustering}, previously discussed. Perhaps, to be expected if we consider that the vast majority of weights actually lie in the ``bulk'' of the scatter plots, while the linear behaviour is only evident for systems with extremely large weights.

\subsubsection{Symbolic Regression}
Whilst NNs have limited interpretability due to the large number of constituent functions being concatenated, there are other methods of supervised learning that are more directly interpretable for extracting mathematical insight.

With the knowledge that NNs can well predict $h^{1,1}$ values of Calabi-Yau four-folds from the ambient weighted projective space weights alone, there is hence experimental evidence for approximate formulas connecting directly these integers.
Motivated by this, in this section techniques of symbolic regression are implemented via the \texttt{gplearn} library to search for candidate approximation formulas.

\begin{figure}[tb]
    \centering
    \includegraphics[width=0.6\textwidth]{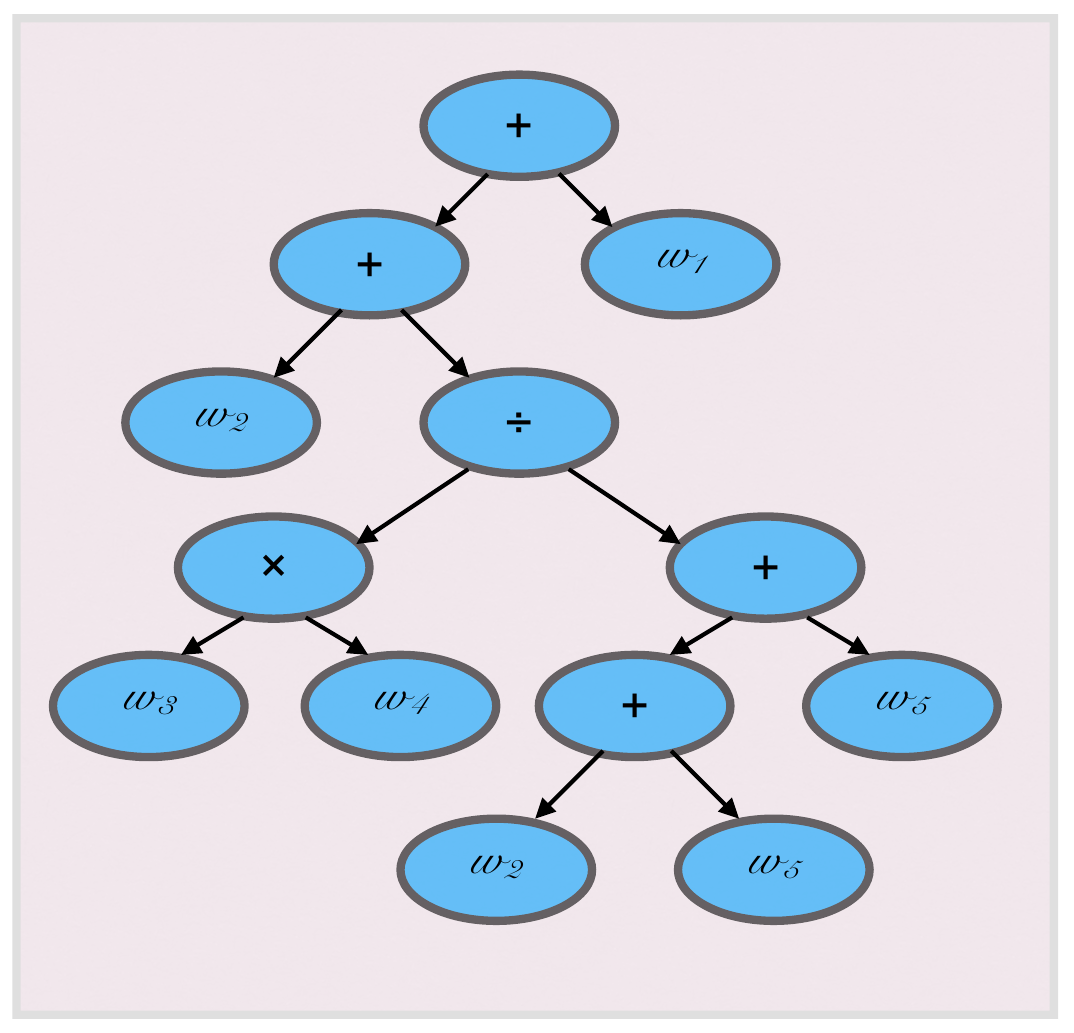}
    \caption{An expression tree representing a candidate learnt formula: $w_1+w_2+\frac{w_3w_4}{w_2+2w_5}$, via symbolic regression.}\label{fig:expressiontree}
\end{figure}

Symbolic regression is a method of supervised learning implemented via a genetic algorithm. 
Initially a basis of functions is provided to the agent, here we will restrict ourselves to the standard normal division algebra basis: $\{+, -, \times, \div\}$.
Then a population of candidate expressions is randomly initialised as expression trees; where expression trees diagrammatically represent formulas as demonstrated in Figure \ref{fig:expressiontree}.
The population of expressions is then evaluated on the training data, noting a parsimony factor rewarding simpler expressions, and many of the best performing expressions are selected for breeding by swapping randomly selected subtrees.
The output of the breeding is a new population of expressions which are then randomly mutated in a variety of ways to produce the next generation.
This process of evaluating, breeding, mutation is then iterated for a fixed number of generations, where the best expression is then selected from the final generation's population as the output.
This output expression is then tested on the test data to produce the final performance measures, here using the same as for the NNs.

After 50 generations of 1000 expressions, with the \texttt{gplearn} recommended breeding and mutation factors and a parsimony of 0.8, the final output candidate expressions as well as performance measures for 3 independent runs were as given in Table \ref{tab:sr}.

\begin{table}[!tb]
\centering
\begin{tabular}{|c|c|c|}
\hline
Expression & $R^2$ & MAPE \\ \hline
$w_1+w_2+\frac{w_3w_4}{w_2+2w_5}$        & 0.884   & 0.322  \\ \hline
$\frac{3}{4}w_1+w_2+\frac{3}{8}w_3$        & 0.872   & 0.332  \\ \hline
$w_1+w_2+\frac{1}{6}w_4$        & 0.860   & 0.339  \\ \hline
\end{tabular}
\caption{Candidate expressions for $h^{1,1}$ as functions of the 6 weights $(w_0,w_1,w_2,w_3,w_4,w_5)$ in the input transverse weight systems from independent symbolic regression runs, with respective performance measures.}
\label{tab:sr}
\end{table}

With the goal of extracting just the first order behaviour for the NN approximate formula, the high parsimony and simple function basis used has limited this regression performance, leading to expected lower performance scores relative to the NNs.
Generalisation to a broader basis with lower parsimony can provide expressions with higher performance, well demonstrated by a run using the full \texttt{gplearn} basis 
\begin{equation}
    \big\{ +, -, \times, \div, -( \cdot ), \sqrt{ \cdot }, \frac{1}{\cdot}, | \cdot |, \text{log}( \cdot ), \text{max}( \cdot ), \text{min}( \cdot ), \text{sin}( \cdot ), \text{cos}( \cdot ), \text{tan}( \cdot ) \big\} \;,
\end{equation}
producing the expression
\begin{align}
    & \sqrt{w_3+3\sqrt{w_4}+\text{min}\bigg(\frac{w_4}{\text{log}(\text{cos}(w_0)-|w_1|)},2w_1 + \text{min}\bigg((w_3-w_0),...}\\
    & \overline{\sqrt{w_2\text{min}\bigg(\sqrt{w_2^2\text{log}(\text{cos}(w_0)-|w_1|)},\text{min}\bigg((w_3-w_0),\text{min}\bigg(\sqrt{w_2w_3},\sqrt{\frac{w_2^2w_5}{w_3-w_0}}\bigg)\bigg)\bigg)}\bigg)\bigg)} \;,\nonumber
\end{align}
with $R^2$ score $0.896$.
However due to the far higher equation complexity and this relatively minimal increase in performance, the lack of interpretability puts motivation on consideration of the initially specified simple basis, with candidate expressions quoted in Table \ref{tab:sr}.
Where these three independent expressions have similar performance, and some similar structure.

The first thing to note is that in each equation there are three summed terms, which are each positive functions of weights.
More specifically, each has a term equal to $w_2$, and another term either equal or proportional to $w_1$.
The occurrence of these earlier weights somewhat corroborates the importance of earlier parts of the weight system seen in §\ref{sec:h11saliency}, however without the $w_0$ factor -- which may be related to $w_0$ having a significantly smaller range.
In each case there is one further term involving higher weights, and across these expressions all additional weights do occur in this term.
Overall, it is quite surprising how well such simple expressions can perform at predicting the $h^{1,1}$ values, and the simple linear sum behaviour does support there being an approximate linear relationship as observed in Figure \ref{fig:h11R_vs_nR}.

\subsection{Classifying CY Property}\label{sec:ml_classify}
The generation of weight system subdatasets for each property combination, as described in §\ref{sec:Partitioning}, enables the design of ML experiments to distinguish these properties in weight systems.
In these cases, the problem is setup as supervised classification, again using the same NN architecture throughout these subinvestigations for consistency and ease of comparison\footnote{We note that tuning hyperparameters leads to improved learning performance, however here we are only focused on showing the existence of good learning, maintaining consistent architecture hyperparameters to compare between investigations.}.

To investigate the stability of the partition, a multiclassification investigation is carried out between all 8 subdatasets.
Subsequently, a binary classification investigation is then carried out to probe the ability of ML architectures to identify each considered property: IP, Intradivisibility, Reflexivity, Transversality (i.e. CY); for each of these the datasets in each of the 2 classes were formed by taking appropriate unions of the partition subdatasets.
To avoid problems caused by unbalanced datasets, during training class weights were fed into the NN such that it is proportionally more rewarded for correctly classifying weight systems in smaller classes; furthermore the MCC performance measure was used, which is known to be unaffected by unbalanced class sizes -- in this sense the MCC is the more appropriate measure of learning.

The investigations, with the appropriate partitions of the partition subdatasets, as well as class sizes, and finally the averaged learning results over the 5-fold cross-validation are presented in Table \ref{tab:classification_results}.
\begin{table}[!t]
\centering
\addtolength{\leftskip}{-1.75cm}
\addtolength{\rightskip}{-1.75cm}
\begin{tabular}{|c|c|c|c|c|}
\hline
Investigation        & Data Partition            & Class Sizes                                                & Accuracy                                                     & MCC                                                         \\ \hline
Multiclassification  &  \begin{tabular}[c]{@{}c@{}} $\{0\}, \{1\}, \{2\}, \{3\}$, \\ $\{4\}, \{5\}, \{6\}, \{7\}$ \end{tabular}                        & \begin{tabular}[c]{@{}c@{}}[408124, 9614, 999975, 988436,\\ 172462, 81215, 847122, 252933]\end{tabular} & \begin{tabular}[c]{@{}c@{}}0.796 \\ $\pm$ 0.007\end{tabular} & \begin{tabular}[c]{@{}c@{}}0.740\\ $\pm$ 0.009\end{tabular} \\ \hline
IP                   & $\{0,1\},\{2,3,4,5,6,7\}$ & [417738, 3342143]                                          & \begin{tabular}[c]{@{}c@{}}0.963\\ $\pm$ 0.001\end{tabular}  & \begin{tabular}[c]{@{}c@{}}0.808\\ $\pm$ 0.008\end{tabular} \\ \hline
Intradivisible        & $\{0,2,3\},\{1,4,5,6,7\}$ & [2396535, 1363346]                                        & \begin{tabular}[c]{@{}c@{}}0.906\\ $\pm$ 0.001\end{tabular}  & \begin{tabular}[c]{@{}c@{}}0.795\\ $\pm$ 0.003\end{tabular} \\ \hline
Reflexive            & $\{2,4,6\},\{3,5,7\}$      & [2019559, 1322584]                                         & \begin{tabular}[c]{@{}c@{}}0.848\\ $\pm$ 0.002\end{tabular}  & \begin{tabular}[c]{@{}c@{}}0.681\\ $\pm$ 0.003\end{tabular} \\ \hline
Calabi-Yau           & $\{0,1,2,3,4,5\},\{6,7\}$ & [2659826, 1100055]                                         & \begin{tabular}[c]{@{}c@{}}0.940\\ $\pm$ 0.002\end{tabular}  & \begin{tabular}[c]{@{}c@{}}0.852\\ $\pm$ 0.002\end{tabular} \\ \hline
Calabi-Yau Reflexive & $\{6\},\{7\}$             & [252933, 847122]                                           & \begin{tabular}[c]{@{}c@{}}0.774\\ $\pm$ 0.001\end{tabular}  & \begin{tabular}[c]{@{}c@{}}0.132\\ $\pm$ 0.009\end{tabular} \\ \hline
\end{tabular}
\caption{Classification results for various partitions of the weight system data. The table shows the mean Accuracy and MCC scores, to 3 decimal places, with standard error, across the 5 cross-validation runs, for the respective investigations labelled by the property being distinguished. The first investigation is multiclassification between all 8 partitions of the weight system data: \{CnIPnD, DnIP, IPnRnD, IPRnD, DnR, DR, CYnR, CYR\}, the remaining investigations are binary classifications between unions of these non-overlapping datasets as labelled by the index in the stated list of weight system partitions. The class sizes are also given for reference (where the second class exhibits the investigated property), many are approximately balanced classifications but where they are not the MCC is a more appropriate non-bias measure.}\label{tab:classification_results}
\end{table}

These classification results are all considerably strong.
For the multiclassification problem, an untrained NN would have null performance expressed by an accuracy $\sim 0.125$ and MCC $\sim 0$, however both performance measures are substantially higher than these scores.
Therefore despite the weight systems being generally indistinguishable by eye, the NNs can learn to extract the appropriate property information sufficiently enough to classify well. 
Examining further the classification output, the averaged normalised confusion matrix for this multiclassification investigation is given by
\begin{equation}
    \begin{pmatrix} 0.084 & 0.000 & 0.012 & 0.004 & 0.000 & 0.000 & 0.009 & 0.000 \\ 0.000 & 0.000 & 0.000 & 0.001 & 0.001 & 0.000 & 0.000 & 0.000 \\ 0.003 & 0.000 & 0.263 & 0.000 & 0.000 & 0.000 & 0.000 & 0.000 \\ 0.001 & 0.000 & 0.000 & 0.249 & 0.006 & 0.000 & 0.005 & 0.000 \\ 0.000 & 0.000 & 0.000 & 0.032 & 0.012 & 0.000 & 0.001 & 0.000 \\ 0.000 & 0.000 & 0.000 & 0.015 & 0.006 & 0.000 & 0.001 & 0.000 \\ 0.009 & 0.000 & 0.001 & 0.024 & 0.005 & 0.000 & 0.185 & 0.002 \\ 0.003 & 0.000 & 0.000 & 0.007 & 0.001 & 0.000 & 0.054 & 0.002 \end{pmatrix}\;,
\end{equation}
to 3 decimal places, where the row is the true class and column the predicted class.
The matrix diagonal represents correctly classified weight systems, and as can be seen the NNs prioritise the first, third, fourth, and sixth classes such that the coprime weight systems with none of the properties are well distinguished from IP and CY subdatasets -- these are also the most populous classes.
The off-diagonal terms are mostly zero, indicating good learning.
However, the demands of this multiclassification problem are high, the architectures must learn to identify many quite different properties simultaneously.
Despite this, the surprising success motivates the binary classification of each property individually.

Each of the binary classification investigations exhibits higher performance measures than multiclassification, indicating that the architectures unsurprisingly perform better when learning one weight system property at a time.
Theoretically these trained NNs could then each be used in turn to identify the properties of a new candidate weight system, and which part of the partition it probably lies in.
The benefit of this is the computation of, particularly, IP and reflexivity becomes especially expensive for larger weight systems where the respective polytope is large and then calculating the dual polytope to check these properties takes increasingly more memory and time.
With the trained NNs, candidate weight systems could be fed into these NNs allowing quick elimination of weight systems unlikely to satisfy these properties.
Then the expensive analytic checks can be performed for the filtered weight system database, producing a far higher proportion of weight systems with the desired properties.

Focusing on the MCC scores, the architectures struggle most with identifying reflexivity, especially in the CY case.
Considering the number of steps required to compute this analytically, via construction of the respective lattice polytope, taking the dual polytope, and then performing many integer checks of the corresponding vertices, this is perhaps not surprising.
Conversely, the performance for identifying the IP property is then surprisingly high, which still requires generating the polytope.
Therefore it is likely the NNs can approximate the polytopes within their architectures but struggle with the integer checks of vertices -- a property notoriously evasive for ML \cite{testolin2023neural}.
Respectively, the NNs can well learn the intradivisibility property, a more direct computation with the weight data, however still with a number of necessary checks.
Finally, and most pleasingly, the CY property can be well learnt for four-fold weight systems, a result observed for three-folds in \cite{Berman_2022}. 
The ability to so successfully predict the existence of further singularity structure in the respective hypersurfaces beyond that of the ambient weighted projective space remains astounding, for a method which is still unclear how to perform directly without using the Landau-Ginzburg string interpretation.

\subsubsection{NN Gradient Saliency}
\begin{figure}[tb]
    \centering
    \begin{subfigure}{0.47\textwidth}
        \centering
        \includegraphics[width=0.98\textwidth]{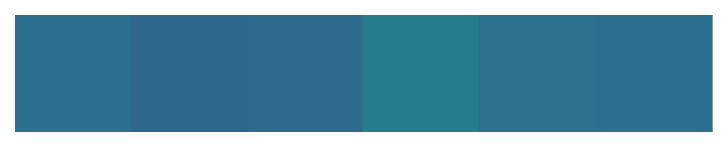}
        \caption{Multiclassification}
    \end{subfigure} 
    \begin{subfigure}{0.47\textwidth}
        \centering
        \includegraphics[width=0.98\textwidth]{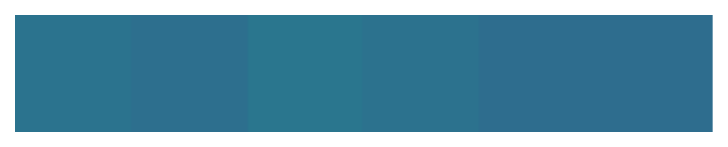}
        \caption{IP}
    \end{subfigure} \\
    \begin{subfigure}{0.47\textwidth}
        \centering
        \includegraphics[width=0.98\textwidth]{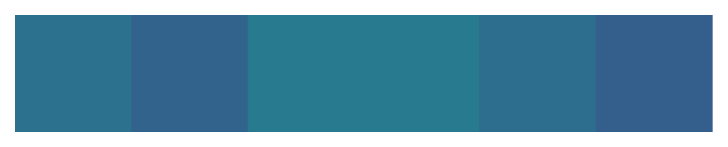}
        \caption{Intradivisible}
    \end{subfigure} 
    \begin{subfigure}{0.47\textwidth}
        \centering
        \includegraphics[width=0.98\textwidth]{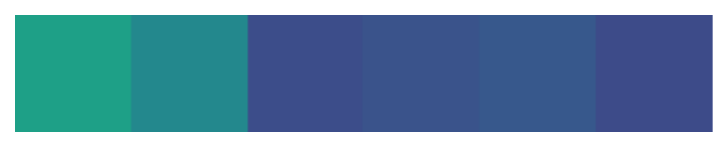}
        \caption{Reflexive}
    \end{subfigure} \\
    \begin{subfigure}{0.47\textwidth}
        \centering
        \includegraphics[width=0.98\textwidth]{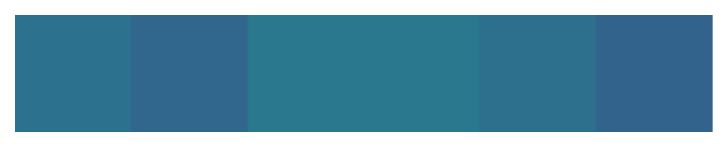}
        \caption{Calabi-Yau}
    \end{subfigure} 
    \begin{subfigure}{0.47\textwidth}
        \centering
        \includegraphics[width=0.98\textwidth]{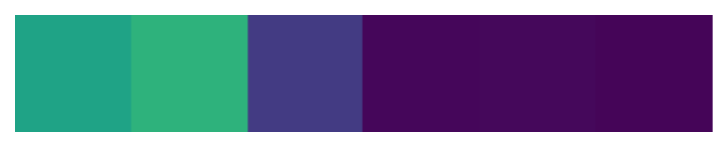}
        \caption{Calabi-Yau Reflexive}
    \end{subfigure} 
    \caption{NN gradient saliency scores for the property classification supervised learning on input weight systems. The lighter colours indicate a larger normalised absolute gradient for that weight in the input 6-vector weight systems, where the saliency scores are averaged over the full test sets of each investigation and each of the 100 repetitions of the investigations.}\label{fig:NNSaliency}
\end{figure}

The relative saliency values for each of the classification tasks are represented in Figure \ref{fig:NNSaliency}.
For these investigations the saliency scores only show significant dependence on the input features for the reflexivity identification, where the earlier weights in the sorted weight systems (and hence smaller weights) are more important in determining the classification.
This is accentuated for the CY reflexivity investigation.
This is likely related to the distribution in Figure \ref{fig:h11R_vs_nR}, where reflexive weight systems appear to be skewed towards lower weights.

%%%%%%%%%%%%%%%%%%%%%%%%%%%%%%%%%%%%%%%%%%%%%%%%%
\section{The Approximation}
\label{sec:Approximation}
The computation of the Hodge numbers for Calabi-Yau four-folds as hypersurfaces in weighted projective spaces was performed in \cite{Lynker_1999} via the Landau-Ginzburg model. Such a calculation involves constructing a number of Poincar\'e-type polynomials and summing their contributions. To do so, one has to perform polynomial multiplications and divisions, which become computationally expensive when the sum of the weights takes large values. This is also the regime where the linear clustering behaviour is more manifest. In the present section, we introduce an approximation for the Hodge numbers, which is well-defined for all Calabi-Yau weight systems, always provides a lower bound, and is significantly faster to compute. 
\subsection{Presenting the Formula}
We first review the standard calculation by following \cite{Batyrev:2020ych}. 
Given a weight system $(w_0, \cdots, w_n)$, we define:
\begin{align}
\left(q_0, q_1, \ldots, q_n\right)=\left(\frac{w_0}{w}, \frac{w_1}{w}, \ldots, \frac{w_n}{w}\right) \in \mathbb{Q}_{>0}^{n+1} \, .
\end{align}
Moreover, for $0 \leq l < w$, we further define:
\begin{equation}
\begin{split}
\theta(l)&=\left(\theta_0(l), \theta_1(l), \ldots, \theta_n(l)\right)=\left(l q_0, l q_1, \ldots, l q_n\right)\, ,\\
\widetilde{\theta}(l)&=\left(\widetilde{\theta}_0(l), \widetilde{\theta}_1(l), \ldots, \widetilde{\theta}_n(l)\right) \in[0,1)^{n+1} \, ,
\end{split}
\end{equation}
with $\widetilde{\theta}$ being the canonical representative of $\theta(l)$ in $(\mathbb{R} / \mathbb{Z})^{n+1}$. To conclude, we reintroduce the last two quantities that appear in the formula for the Hodge numbers:
\begin{equation}
\begin{split}
\operatorname{age}(l)&=\sum_{i=0}^n \widetilde{\theta}_i(l)=\sum_{\widetilde{\theta}_i(l) \neq 0} \widetilde{\theta}_i(l)   \, , \\
\operatorname{size}(l)&=\operatorname{age}(l)+\operatorname{age}(w_{tot}-l) \, .
\end{split}
\end{equation}
Given these ingredients, then the full formula for the Hodge numbers $h^{p,q}$ reads:
\begin{align}
    \sum_{p, q}(-1)^{p+q} h^{p, q} u^p v^q = \frac{1}{u v} \sum_{0 \leq l<w_{tot}}\left[\prod_{\tilde{\theta}_i(l)=0} \frac{(u v)^{q_i}-u v}{1-(u v)^{q_i}}\right]_{i n t} (-u)^{\operatorname{size}(l)}\left(\frac{v}{u}\right)^{\operatorname{age}(l)} \, .
    \label{eq:Exact_formula}
\end{align}
It is evident that the polynomial products and divisions within the square brackets are what takes most of the computational resources. Just for reference, we note that for high weights, the software \texttt{sagemath} cannot perform the calculation due to the high number of terms. For this reason, the algorithm had to be hardcoded directly. One way to go about simplifying this formula is to identify for which values of $l$ the terms in the sum contribute the most. Just by empirical observation, we find that the main contributions come from the element of zero age and the ones with maximal size\footnote{This is well exemplified by looking at examples 4.5 and 4.6 in \cite{Batyrev:2020ych}.}. As we are about to argue in detail, it turns out that including only terms of these types provides a very efficient approximation for the Hodge numbers. It is both very accurate and much faster to implement. Moreover, it bounds the exact results from below. Explicitly, the approximated Hodge numbers $h_A^{p, q}$ can be computed as:
\begin{equation}
\begin{split}
     \sum_{p, q}(-1)^{p+q} h_A^{p, q} u^p v^q = \frac{1}{u v} \Bigg\{ \left[\prod_{i} \frac{(u v)^{q_i}-u v}{1-(u v)^{q_i}}\right]_{i n t} +  \sum_{\mathrm{size}(l) = n} (-u)^{w_{tot}}\left(\frac{v}{u}\right)^{\operatorname{age}(l)} \Bigg \} \, .
     \label{eq:Approx_formula}
\end{split}
\end{equation}
We tested this approximation against the two relevant datasets: the Calabi-Yau four-folds considered in this article and the smaller set of Calabi-Yau three-folds. We start by presenting our findings for the latter case\footnote{For completeness, we point out that the approximation also fails to be well-defined for non-Calabi-Yau weight systems in general.}.

\subsection{Application to Calabi-Yau Three-folds}
\label{sec:Appl_three}
\begin{figure}[tb]\label{fig:R_vs_nR}
\centering
\begin{subfigure}{0.47\textwidth}
\centering
\includegraphics[scale=0.41]{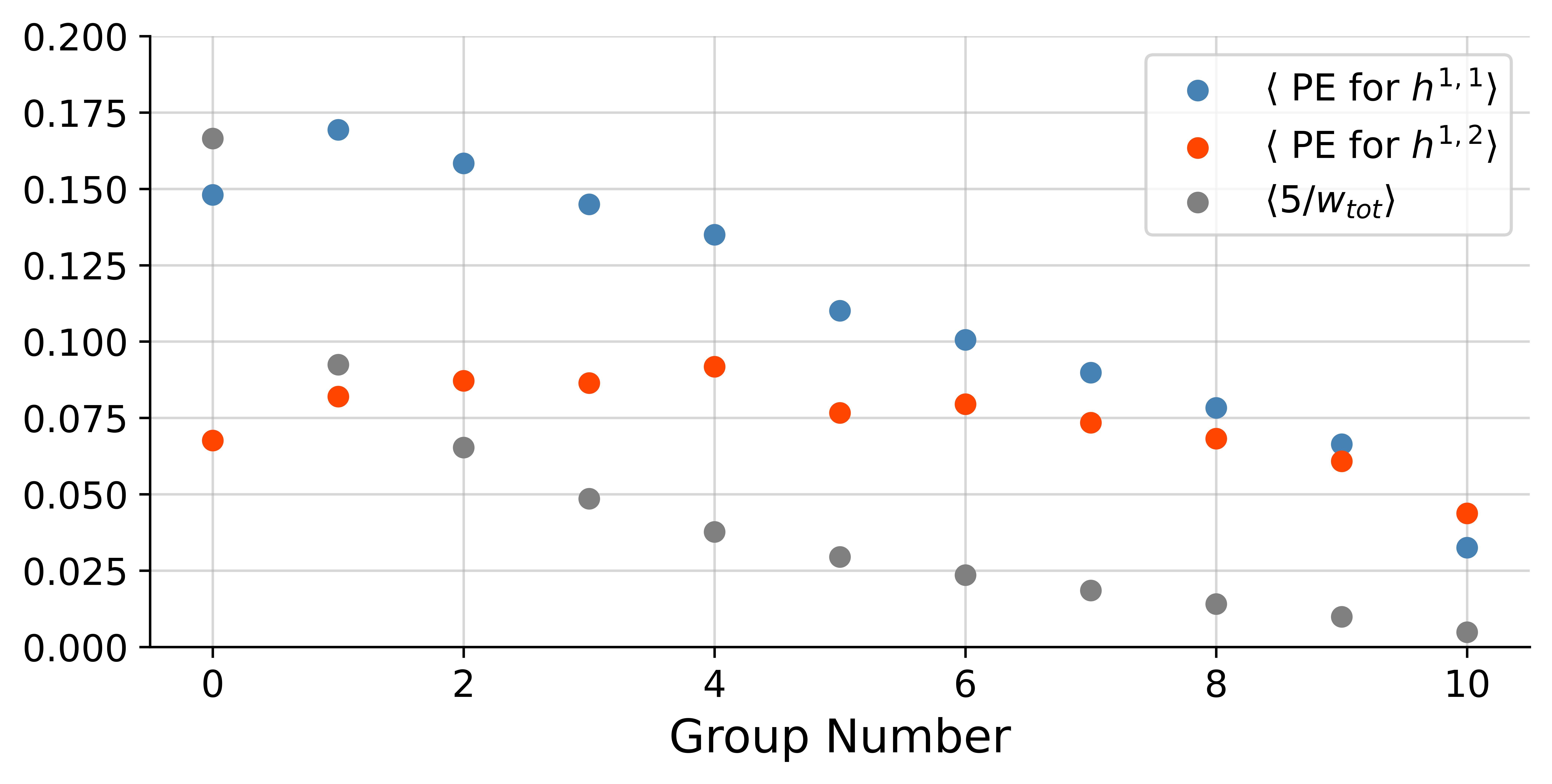}
\caption{}\label{fig:Threefolds_appa}
\end{subfigure}
\begin{subfigure}{0.47\textwidth}
\centering
\includegraphics[scale=0.41]{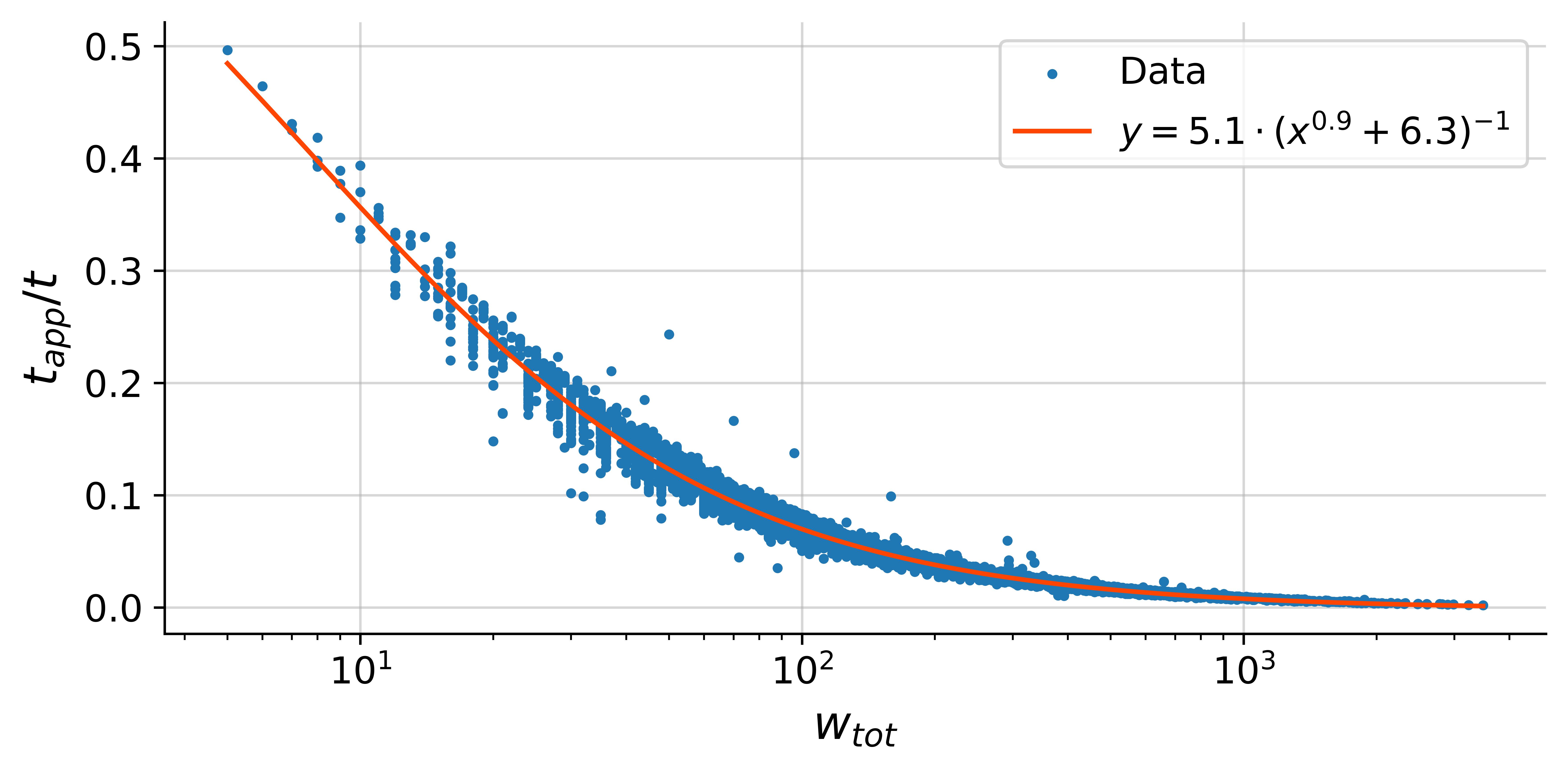}
\caption{}\label{fig:Threefolds_appb}
\end{subfigure}
\caption{These plots summarise the main features of the approximation \eqref{eq:Approx_formula}, both in terms of accuracy and in terms of computational efficiency, compared with the exact formula \eqref{eq:Exact_formula}. The data refer to Calabi-Yau three-folds as hypersurfaces in weighted projective spaces.}
\label{fig:Threefolds_app}
\end{figure}

As just mentioned, \eqref{eq:Exact_formula} becomes more and more involved to compute as the weights in the system become larger. Since the dataset for Calabi-Yau manifolds in weighted projective spaces are already ordered according to the sum of the weights, we conveniently divided the $7555$ three-folds' weight systems (of 5 weights) into $11$ groups, from lowest to highest. The plot in Figure \ref{fig:Threefolds_appa} shows the mean percentage error of the approximation formula \eqref{eq:Approx_formula} for each of those groups, both for $h^{1,1}$ and $h^{1,2}$, showing that the approximation becomes more precise as we go to higher weights. For reference, we also include a measure for the mean sum of weights in each of the groups $\langle5/w_{tot} \rangle $. We observe that the average percentage error for large weights is remarkably small, lying somewhere between $3 \%$ and $5 \%$ for both Hodge numbers. 
\graphicspath{ {./Figures/} }

The plot of Figure \ref{fig:Threefolds_appb} shows the ratio of the computational time against the sum of weights $w_{tot}$. As anticipated, the computational time needed to evaluate \eqref{eq:Approx_formula} is considerably smaller than the time taken by the full version \eqref{eq:Exact_formula}. In fact, their ratio gets to values in the order of $10^{-2}$ for the largest weight systems in the dataset\footnote{One might argue that our implementation of the formula \eqref{eq:Exact_formula} could be further optimised, reducing the time needed to compute the Hodge numbers exactly. However, such an optimisation would lead to a quicker implementation of \eqref{eq:Approx_formula} as well, so that we do not expect the ratio to change significantly.}.

Some other useful figures for this approximation are reported in Table \ref{tab:Threefold_data}.
\begin{table}[tb]
\centering
\begin{tabular}{|c|c|c|}
\hline
\, & $h^{1,1}$ & $h^{1,2}$ \\
\hline
$R^2$ Score  & 0.969  &  0.981 \\
\hline
MAPE  & 0.113  &  0.075  \\
\hline
MAE  & 7.1   &  3.3  \\
\hline
Exact Results  & 32 \%  &  56 \% \\
\hline
\end{tabular}
\caption{This table reports performance measures for the approximation applied to Calabi-Yau three-folds. Specifically, the $R^2$ score, the mean absolute percentage error, the mean absolute error, and the percentage of exact results where the approximation matched the true value.}
\label{tab:Threefold_data}
\end{table}
These results show that the approximation, even though it excludes the vast majority of the terms that appear in \eqref{eq:Exact_formula}, is still able to match the exact values a significant number of times. Moreover, we find another crucial feature of the truncated sum \eqref{eq:Approx_formula}:
\begin{align}
    h_A^{1,1} \leq h^{1,1} \quad \quad \mathrm{and} \quad \quad h_A^{1,2} \leq h^{1,2} \, .
    \label{eq:Lower_bound_threefolds}
\end{align}
Thus, since this is also the case for four-folds, the approximation presented in this work offers a quickly accessible tool for extracting tight lower bounds of the Hodge numbers. 

\begin{figure}[tb]\label{fig:Comp_hist_3}
\centering
\includegraphics[scale=0.6]{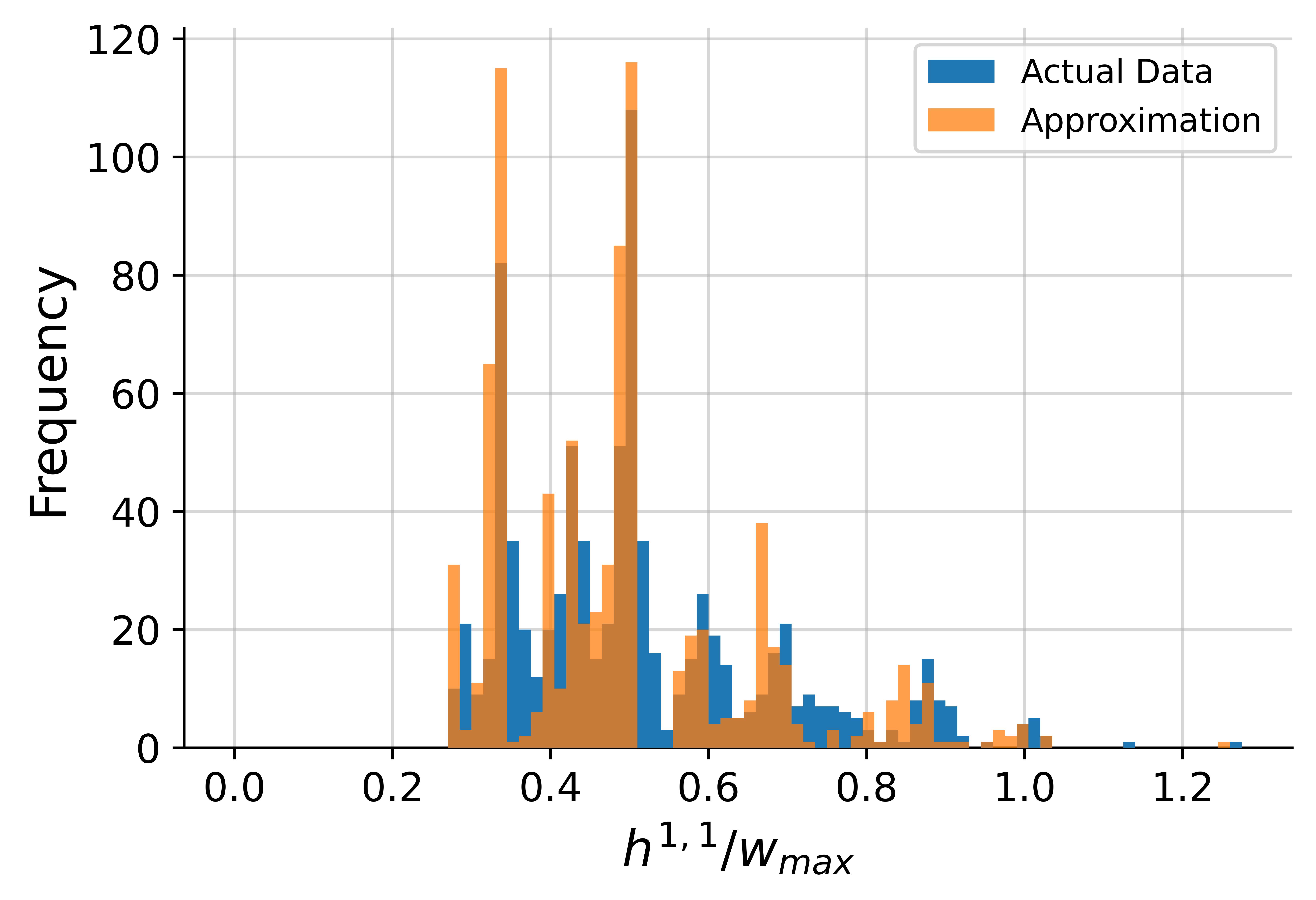}
\caption{This histogram plot shows that the approximated Hodge numbers also cluster for certain values of the ration $h^{1,1}/w_{max}$. Only weight systems with $w_{max}>250$ are plotted here, since this corresponds to the regime where the forking behaviour is most visible. This plot should be compared with the one appearing in \cite{Berman_2022}.}
\label{fig:Threefolds_hist}
\end{figure}

As a final feature, we note that the dataset built from the approximated Hodge numbers $h_A^{1,1/2}$ correctly reproduces the clustering behaviour observed in \cite{Berman_2022}. This is best shown with a histogram plot, in Figure \ref{fig:Threefolds_hist}, where we can clearly see various peaks in $h^{1,1}/w_{max}$, corresponding to the slopes of the clustering lines. They overlap almost completely, showing that the clusters are essentially the same for both datasets: the exact Hodge numbers and the ones obtained via our approximation. Consistently with \eqref{eq:Lower_bound_threefolds}, the peaks associated to $h_A^{1,1}$ are slightly shifted to the left.
\graphicspath{ {./Figures/} }
Thus, by narrowing down the full formula from the Landau-Ginzburg model \eqref{eq:Exact_formula} to a small number of terms, we obtained a much simpler expression, which still reproduces the same behaviour for large weights. This might be a step forward towards the understanding of the linear clustering that characterises the cohomological numbers of Calabi-Yau's in weighted projective spaces.

\subsection{Application to Calabi-Yau Four-folds}
\label{sec:Appl_four}
We now move to the case of four-folds in weighted projective spaces. This dataset is considerably bigger compared to three-folds, with $1100055$ spaces, and it contains systems with very large weights. A natural consequence is that the computational times are much longer, which makes the advantages of the approximation even more evident. To give a concrete example, we focused on the millionth weight system in the dataset, which reads $[  45,    74,  2460, 12792, 17876, 33173]$. Our implementation of \eqref{eq:Exact_formula} takes roughly $40$ hours, while the approximated version \eqref{eq:Approx_formula} is computed in $49$ seconds. The ratio between the two is $0.00034$. Moreover, the approximated results are very accurate: the exact ones are $(h^{1,1} = 10718, \, h^{1,2}=0, \, h^{1,3} = 986, \, h^{2,2} = 46860)$, while the approximated ones read $(10683, \, 0, \, 986 ,  \, 46500)$\footnote{To make this comparison, we had to match our conventions with the ones used for the existing datasets; this is discussed in the next section, after equation \eqref{Hodgecons}.}.

\begin{figure}[tb]\label{fig:Comp_hist_4}
\centering
\begin{subfigure}{0.47\textwidth}
\centering
\includegraphics[scale=0.41]{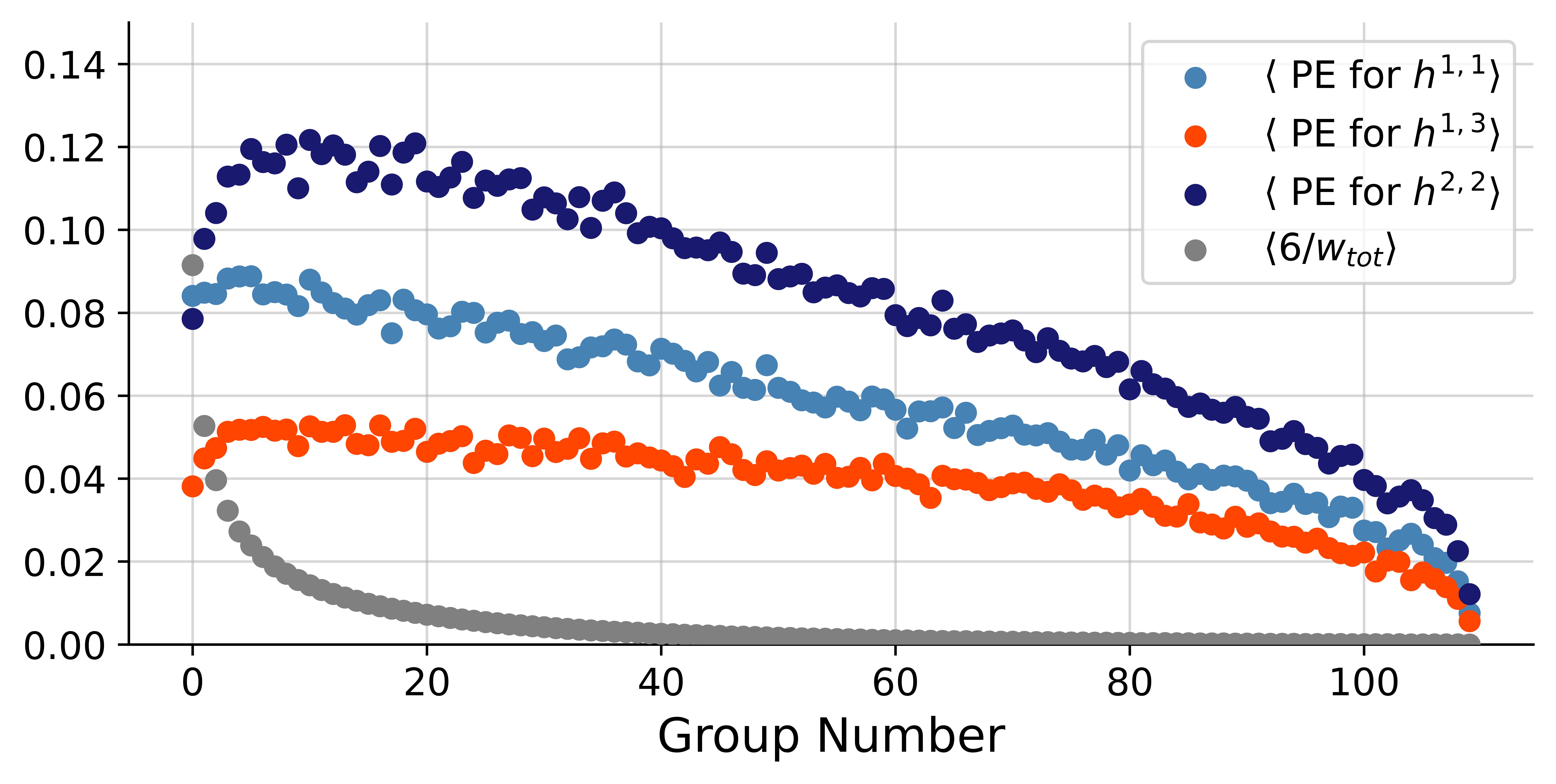}
\caption{}
\label{fig:Fourfolds_approxa}
\end{subfigure}
\begin{subfigure}{0.47\textwidth}
\centering
\includegraphics[scale=0.41]{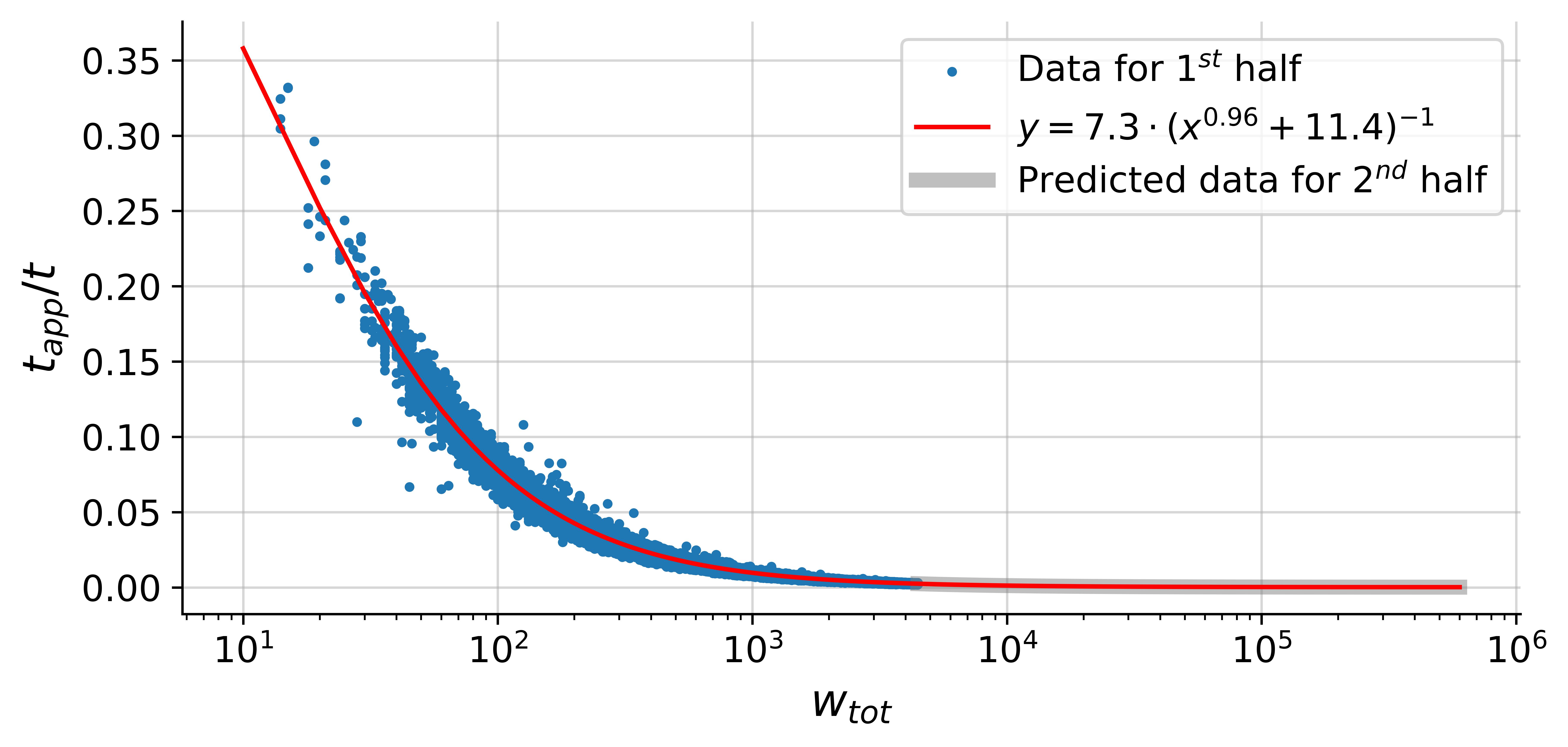}
\caption{}\label{fig:Fourfolds_approxb}
\end{subfigure}
\caption{These plots summarise the main features of the approximation \eqref{eq:Approx_formula}, applied to the dataset of Calabi-Yau four-folds as hypersurfaces in weighted projective spaces. The left plot (a) shows the mean accuracy for groups of $2000$ weight systems ordered according to their sum of weights. We used $20\%$ of the dataset for this plot, sampled uniformly. The computational efficiency is analysed in the right plot (b), where the computational time of the approximation is compared to the one associated with the exact formula \eqref{eq:Exact_formula}. We chose roughly $50000$ samples randomly from the first half of the dataset (blue dots), and used these data to extrapolate the behaviour for the second half. As discussed in the text, the best fit function shown predicts very accurately the ratios for systems with larger weights.}
\label{fig:Fourfolds_approx}
\end{figure}

Regarding the precision of the approximation, we illustrate it in Figure \ref{fig:Fourfolds_approxa}, with a similar plot to the one used for three-folds. We omit $h^{1,2}$ from the picture because it can take the value of zero, making the percentage error not well-defined. We sampled randomly and uniformly $20\%$ of the dataset, and then divided it into groups of $2000$ samples. As before, all samples were ordered according to the sum of weights, then dividing into the groups, as shown by the gray points. We observe once again that the percentage error gets smaller as the weights become larger, reaching roughly $1 \%$ for the samples with largest weights, for all three Hodge numbers.
\graphicspath{ {./Figures/} }
The plot in Figure \ref{fig:Fourfolds_approxb}, on the other hand, shows the comparison between the computational resources employed by the exact expression from the Landau-Ginzburg model and by our approximation. It is evident from the example just discussed that systems with large weights necessitate a very long computation time for the full formula \eqref{eq:Exact_formula}. Thus, we collected data within the first half of the dataset, and then extrapolated our findings to the second half, containing very large weights. The red curve provides a good interpolation of the data, and it turns out to give very accurate predictions as well. This can be confirmed by plugging $w_{tot} = 66420$, which corresponds to the example weight system discussed above, into the expression for the best fit function. The result is $t_{app} / t = 0.00035$, which is remarkably close to the actual ratio ($0.00034$) obtained from the explicit computation. 

Finally, let us report the main properties of the approximation. They are shown in Table \ref{tab:Four_folds}, and they are extracted from the same data plotted in Figure \ref{fig:Fourfolds_approxa}, i.e. from a set of roughly $220000$ random Calabi-Yau weight systems.
\begin{table}[tb]
    \centering
    \begin{tabular}{|c|c|c|c|}
       \hline  & \quad $h^{1,1}$ \quad & \quad $h^{1,3}$ \quad & \quad $h^{2,2}$ \quad \\
        \hline $R^2$ & $0.999$ &$0.999$ & $0.999$\\
        \hline MAPE & $0.058$ &  $0.039$ & $0.082$ \\
        \hline MAE & $74.7$ & $32.4$ & $681.9$ \\
        \hline Exact Results & $26.9 \%$ & $48.7 \% $ & $7.1 \%$\\
        \hline
    \end{tabular}
    \caption{This table reports performance measures for the approximation applied to Calabi-Yau four-folds. Specifically, the $R^2$ score, mean absolute percentage error, the mean absolute error, and the percentage of exact results where the approximation matched the true value.}
    \label{tab:Four_folds}
\end{table}
We end this section by making a final remark. The first one is that, analogously to the three-folds case, the approximation provides a lower bound also for four-folds. However, this is trivially satisfied for $h^{1,2}$, since our approximation always yields zero for this case. Summarising, we have that:
\begin{align}
    h_A^{1,1} \leq h^{1,1} \quad \quad \quad  h_A^{1,2} \equiv 0 \leq h^{1,2} \, , \quad \quad  \quad h_A^{1,3} \leq h^{1,3} \, ,\quad \quad  \quad h_A^{2,2} \leq h^{2,2} \, .
    \label{eq:Lower_bound_fourfolds}
\end{align}

Therefore use of this approximation in practical computations of Hodge numbers not only provides a significant speed improvement, but also will always be a lower bound.
Therefore, in designing string effective theories where the topology of the chosen Calabi-Yau manifold for compactification intrinsically sets many properties of the resulting theory, this approximation allows for incompatible manifolds to be confidently and quickly discarded where any $h^{p,q}_A$ is larger than the desired values for the desired theory being built. 
Additionally, it also provides a good approximation for the remaining candidates, allowing them to be sorted prior to search with the full formula, such that many less manifolds will need to be checked before finding the correct topology for the desired theory.

\section{Higher Weight systems}\label{sec:higherweights}
The approximation has been tested on both three-folds and four-folds. These are the only two existing datasets of Calabi-Yau manifolds built as hypersurfaces in weighted projective spaces. Here, we present the first efforts towards the understanding of these spaces in higher dimensions, by generating a partial dataset of candidate transverse weight systems of 7 weights, and the respective Calabi-Yau five-folds' Hodge numbers. We discuss how the approximation can be used to quickly extract information about such a dataset, and we additionally generate a first partial dataset of candidate transverse weight systems of 8 weights, then using \eqref{eq:Approx_formula} to construct an approximated list of six-folds' Hodge numbers.

\subsection{Calabi-Yau Five-folds}
Calabi-Yau five-folds appear in a number of dimensional reductions in the literature. For instance, it was found that M-theory compactified on a Calabi-Yau five-fold results in an exotic $\mathcal{N}=2$ supersymmetric quantum mechanics \cite{Haupt:2008nu}. 
Moreover, Calabi-Yau five-folds play a role in F-theory, where upon compactification, they provide a way to systematically construct $N = (0,2)$ CFTs \cite{Schafer-Nameki:2016cfr, Tian:2020yex}, which may lead to their classification. Therefore, an extended dataset of such Calabi-Yau manifolds would make it possible to explore the landscape of such CFTs. Additionally, in \cite{Curio:1998bv}, a 3-dimensional string vacua with $\mathcal{N}=1$ supersymmetry has been found, which can be interpreted as a compactification of S-theory on a Calabi-Yau five-fold. Despite their role in the construction of low-dimensional theories, examples of five-folds have not been systematically constructed, until the recent effort in \cite{Alawadhi:2023gxa}, which focuses on the CICY construction. Here, we present a second in that direction, i.e. we generate, for the first time, a subset of Calabi-Yau five-folds obtained as hypersurfaces in $\mathbb{P}^6_{\mathrm{w}}$. Specifically, we generate all 7-weight weight systems whose sum of weights $w_{tot} \leq 200$. To efficiently identify those which have the required property to describe a Calabi-Yau (for more details, see §\ref{sec:background}), we employ a two-step approach. We first systematically search all partitions of each sum of weights up to 200 as weight systems, extracting those which are coprime, IP\footnote{Work in \cite{Skarke:1996hq} showed that transverse weight systems are by necessity IP for any size weight system.}, and intradivisible; performed with our code functionality. Then we use the approximation as a tool for establishing the Calabi-Yau property and select all the weight systems that are well-defined with respect to \eqref{eq:Approx_formula}, or equivalently all those such that the polynomial division $\left[\prod_{i} \frac{(u v)^{q_i}-u v}{1-(u v)^{q_i}}\right]$ gives no reminder. The candidate weight systems identified this way were then all checked with respect to the exact formula \eqref{eq:Exact_formula}, and all turned out to be well-defined, yielding the full exact Hodge diamond. To provide confidence in the generated data, two non-trivial checks on the cohomological data were performed. First, computing the Euler number from the weights alone with \eqref{hodgeeulerformulas}, and then verifying it agrees with the identity 
\begin{equation}\begin{split}
\chi =2h^{1,1}-4h^{1,2}
 +4h^{1,3}
 +2h^{2,2}-2h^{1,4}-2h^{2,3}\; ,
 \end{split}
 \label{Euler0}
\end{equation}
in terms of the computed Hodge numbers.
Moreover, we also checked that the Hodge numbers satisfy the constraint derived from the Atiah-Singer index theorem:
\begin{equation}\label{Hodgecons}
 11 h^{1,1}-10h^{1,2}-h^{2,2}+h^{2,3}+10h^{1,3}-11h^{1,4}=0
  \; .
\end{equation}
Both checks were passed by all the weight systems, which describe new Calabi-Yau geometries in complex dimension five. For reference, let us present five examples of such spaces, shown in Table \ref{tab:Examples_5}. 
\begin{table}[t]
    \centering
    \begin{tabular}{|c|c|}
    \hline
       Weight System  & $[h^{1,1},h^{1,2},h^{1,3},h^{1,4},h^{2,2},h^{2,3},\chi]$ \\ \hline \hline [1,1,1,1,1,1,1]
         & $[1667,     0,     0,     1, 18327,     1, 39984]$ \\
         \hline $[ 1,     6,     8,    12,    14,    19, 60]$ & $[3999,
           0,     4,     3, 44022,    26, 96000]$ \\ \hline $[5,    9,    9,   18,   27,   34,   51]$ & $[577,  201,    0,
         12, 5430, 1225, 8736]$ \\ \hline
         $[4,      4,      5,     10,     27,     45,     85]$ & $[5087,      8,      0,     12,  55938,    193, 121608]$
        \\ \hline
         $[25,    25,    25,    25,    28,    32,    40]$ & $[185,
         350,     1,    30,   566,  2351, -4656]$ \\ \hline
    \end{tabular}
    \caption{Examples of weight systems with 7 weights, describing Calabi-Yau five-folds, together with the associated invariants.}
\label{tab:Examples_5}
\end{table}
We point out a small difference in definitions between this section and the previous ones. In §\ref{sec:Appl_three} and §\ref{sec:Appl_four}, we compared our results with the existing datasets (which can be found at \cite{KSdata}), whose conventions are slightly different from ours. Namely, for Calabi-Yau three-folds, $h^{1,1}$ and $h^{1,2}$ determined using \eqref{eq:Exact_formula} have to be exchanged in order to match \cite{KSdata}. Analogously, $h^{1,1}$ and $h^{1,3}$ should be swapped for four-folds to be consistent with the existing list. For the reminder of this paper, we present our results as they are obtained from \eqref{eq:Exact_formula}.

Having mentioned this subtlety, we note that a quick consistency check of our results comes straightforwardly from considering the first entry in Table \ref{tab:Examples_5}. This weighted projective space is trivial (i.e. is not actually weighted), so that it gives rise to the simplest Calabi-Yau five-fold defined by a degree seven polynomial in $\mathbb{P}^6$. The associated cohomology matches the result reported in appendix of \cite{Haupt_2009}. 

The global properties of Calabi-Yau five-folds as hypersurfaces in weighted projective spaces, with sum of weights up to $200$, read:
\begin{align}
&\left\langle h^{1,1}\right\rangle= 9004.6_{\,71}^{ \, 2314879}\,, \quad  \left\langle h^{1,2}\right\rangle=24.0_{\,0}^{15180} \, , \quad  \left\langle h^{1,3}\right\rangle= 2.4_{\, 0}^{\, 703}\, , \quad  \left\langle h^{1,4}\right\rangle=8.8_{\, 1}^{\, 50}\,  \nonumber \\[1em] &\left\langle h^{2,2}\right\rangle=98932.3_{\, 566}^{\, 25463560} \, , \quad \langle h^{2,3} \rangle=194.9_{\, 1}^{\, 64930}
\, , \quad \langle\chi\rangle=215379.9_{\, -13248}^{\, 55556832} \, .
\label{eq:Hodges_fivefolds}
\end{align}
The dataset of the 274730 weight systems with their topological properties is made available on \href{https://github.com/Tancredi-Schettini-Gherardini/P5CY4ML}{GitHub}, presented in the format [[$w_i$], $w_{tot}$, [$h^{p,q}$], $\chi$], where $h^{p,q}$ are written in the same order as in \eqref{eq:Hodges_fivefolds} above. With some further analysis in Figure \ref{fig:Five_folds}.

As it can be guessed by looking at $\left\langle h^{1,1}\right\rangle$ and $\left\langle h^{1,4}\right\rangle$, the subset of spaces considered here does not show mirror symmetry. This is due to the fact that we only restricted ourselves to a small sum of weights, whose mirror-symmetric pairs lie in the large-weights regime. We expect the cohomological data in that regime to be practically inaccessible, due to the large computational times associated with \eqref{eq:Exact_formula}. For this reason, we believe that the approximation presented in §\ref{sec:Approximation}, which proved to be extremely accurate for large weights in the four-folds investigation, could be a key tool for attempting such a task. Moreover, we also expect the list of all possible Calabi-Yau 7-weight weight systems to be astronomical in size. Once again, the truncated formula \eqref{eq:Approx_formula} provides a quickly computable tight \textit{lower bound} for the Hodge numbers of all those yet undiscovered manifolds.

\subsection{Calabi-Yau Six-folds}
While their role in physics is marginal (they could only be employed for compactifications of S-theory), Calabi-Yau six-folds have their own relevance directly within mathematics. These spaces could provide additional information about the - still very mysterious to this day - landscape of Calabi-Yau geometries, as they are the second non-trivial family of Calabi-Yau manifolds in even complex dimensions. Their construction as hypersurfaces of weighted projective spaces involves using 8-weight weight systems, which are both more numerous and effectively infeasible to run bulk computation of exact topological parameters. For these reasons, we find the truncated approximation formula to be especially pertinent, allowing computation of approximated Hodge values for all the generated candidate transverse weight systems with $w_{tot} \leq 200$. Once again, we first identify the IP intradivisible weight systems, and then select the ones which are well-defined with respect to \eqref{eq:Approx_formula}, numbering 1482022 candidate transverse weight systems of 8 weights (accessible at this work's respective \href{https://github.com/Tancredi-Schettini-Gherardini/P5CY4ML}{GitHub} in the same format as for the five-folds). From there the Euler number was also computed exactly with the less computationally-intensive direct formula from weights \eqref{hodgeeulerformulas}. A few examples are reported in Table \ref{tab:Examples_6}.
\begin{table}[t]
    \centering
    \begin{tabular}{|c|c|}
    \hline
       Weight System  & $[h^{1,1}_A,h^{1,2}_A,h^{1,3}_A,h^{1,4}_A,h^{1,5}_A,h^{2,2}_A,h^{2,3}_A,h^{2,4}_A,h^{3,3}_A,\chi]$ \\ \hline \hline [1,1,1,1,1,1,1,1]
         & $[6371,       0,       0,       0,       1,
        154645,       0,       1,  398568, 720608]$ \\
         \hline $[1,         1,         3,         4,        15,        22,
              43,        44 ]$ & $[ 265283,         0,         0,
               0,         6,   6629968,         0,        27,  16974104,
       30764676]$ \\ \hline $[1,         1,         3,         8,        15,        16,
              40,        84]$ & $[484547,         0,         0,
               0,         4,  12217438,         0,        17,  31218692,
       56622576]$ \\ \hline
         $[7,      7,      7,     14,     14,     25,     41,     74]$ & $[9905,      0,      0,      0,     22, 109916,      0,
           40, 207950, -162552]$
        \\ \hline
         $[18,     20,     20,     25,     25,     26,     30,     36]$ & $[344,      0,      0,      0,      6,   7499,      0,
           37,  19303, 34360]$ \\ \hline
    \end{tabular}
    \caption{Examples of weight systems with 8 weights, describing Calabi-Yau six-folds, together with the associated invariants.}
\label{tab:Examples_6}
\end{table}

\graphicspath{ {./Figures/} }
\begin{figure}[t]\label{fig:Multi_CY}
\centering
\includegraphics[scale=0.8]{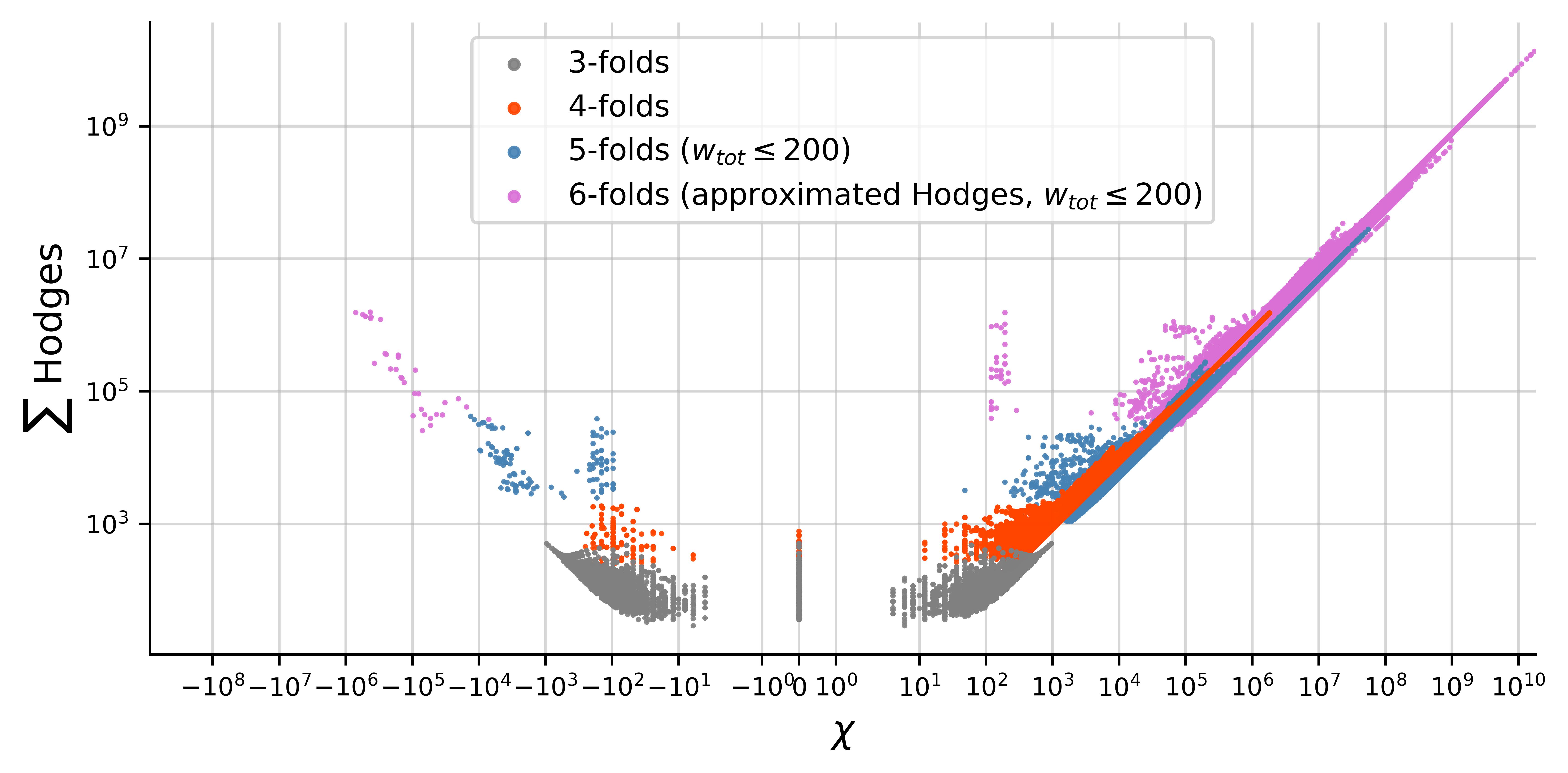}
\caption{This plot illustrates some features of the cohomological data and Euler numbers for Calabi-Yau manifolds constructed as hypersurfaces in weighted projective spaces. The three-folds and four-folds' data points exhaust all possible Calabi-Yau manifolds of that type. For five-folds and six-folds, we restricted ourselves to weight systems with $w_{tot} \leq 200$. These are all newly discovered geometries. The Hodge numbers for six-folds were obtained through the approximation \eqref{eq:Approx_formula}, hence they represent tight lower bounds of these cohomological properties.}
\label{fig:Five_folds}
\end{figure}
Similarly to before, the first manifold in Table \ref{tab:Examples_6} is nothing but the Calabi-Yau six-fold defined by a degree 8 polynomial in $\mathbb{P}^7$. We find that our results, despite coming from the truncated formula, exactly match the exact Hodge numbers, which can be found in \cite{Dumachev2015CompleteIC}.

The preliminary analysis is also shown in Figure \ref{fig:Five_folds}, where a comparison across different complex dimensions is shown; it illustrates that, for a low sum of weights, the five-folds and six-folds are appropriately skewed towards positive Euler numbers. The approximated Hodge numbers show a similar behaviour to what one would expect based on the other distributions. Just for reference, we report here the main global features of the approximated invariants that we computed:
\begin{align}
&\left\langle h^{1,1}_A\right\rangle= 96686.7_{\,279}^{ \,147270231}\,, \quad   h^{1,2}_A \equiv 0 \, , \quad   h^{1,3}_A \equiv 0 \, , \quad  h^{1,4}_A \equiv 0 \, ,  \quad \left\langle h^{1,5}_A \right\rangle=4.3_{\, 1}^{\, 28} \, , \nonumber \\[1em]
&\left\langle h^{2,2}_A \right\rangle=2424722.8_{\, 6366}^{\, 3759686446}
\, , \quad  h^{2,3}_A \equiv 0 \, , \quad \left\langle h^{2,4}_A \right\rangle=27.5_{\, 1}^{\, 65} \, ,  \\[1em]
&\left\langle h^{3,3}_A \right\rangle=6189235.0_{\, 15542}^{\, 9581156426}
\, , \quad  \left\langle \chi \right\rangle=11309730.8_{\, -708480}^{\, 17395069848} \, . \nonumber
\label{eq:Hodges_sixefolds}
\end{align}

%%%%%%%%%%%%%%%%%%%%%%%%%%%%%%%%%%%%%%%%%%%%%%%%%
\section{Summary \& Outlook}\label{sec:conclusion}
This work was focused on, but not limited to, the analysis of Calabi-Yau four-folds obtained as hypersurfaces in weighted projective spaces. By restricting to systems with large weights, a linear clustering behaviour analogous to the one found for three-folds in \cite{Berman_2022} was observed and quantitatively corroborated through the K-Means clustering normalised inertia. By gradually relaxing the conditions on the weights, we were able to produce a partition of coprime weight systems according to the most relevant properties: IP, reflexivity, intradivisibility, and transversality (Calabi-Yau). Generating datasets for each subset in such a partition. 

While all of the above was performed using concrete analytic algorithms, statistical machine learning techniques were also applied both to the dataset of Calabi-Yau four-folds and to the partitioned set of more general weight systems. Regarding the former, a fully connected regressor network was shown to predict the cohomological Hodge data, and the Euler number, from the system weights. We found particularly good results, with $R^2 \sim 0.91$, for $h^{1,1}$, on the whole dataset. For the other invariants, we observed very different results for systems with small weights as opposed to systems with large weights. For instance, $h^{1,3}$ and $h^{2,2}$ showed results with $R^2 > 0.90$ for the half of the dataset containing lower weights, while the accuracy dropped significantly for the other half. These three numbers provide sufficient information to determine the full Hodge diamond, however results were also reported associated with $h^{1,2}$, which had a poor performances since it is zero $48 \%$ of the time, and $\chi$, which showed similar results to $h^{1,3}$ and $h^{2,2}$. 

The partition of weight systems according to their respective properties within \{IP, reflexive, intradivisible, transverse\}, where transversality implied the existence of a Calabi-Yau hypersurface, was classified with the respective fully connected classification architecture.
Multiclassification results were surprisingly high between all parts of the partition, reaching MCC scores of 0.740. 
Separately, binary classification investigations managed to well identify each property respectively from unions of the partition subdatasets, struggling most with reflexivity.

Motivated by the strong performances of the neural networks, and inspired by the interpretability of the gradient saliency analysis and symbolic regression, we explored a simpler truncated version of the formula coming from the Landau-Ginzburg model used for calculating the Calabi-Yau Hodge numbers from the ambient $\mathbb{P}^n_\textbf{w}$'s weight system. This approximation drastically reduces the number of terms involved in the computation, making it easier to study analytically, and substantially faster to compute numerically. Its main features are: it provides a tight \textit{lower} bound for the Hodge numbers; it is especially accurate for systems with large weights (average MAPE of $<1\%$ for the $10000$ systems with largest weights); it is dramatically faster than the exact formula (up to $10^4$ times quicker); it reproduces the observed linear clustering behaviour for large weights. 

Finally, motivated by the speed improvements available from this approximation, transverse weight systems (satisfying the necessary intradivisible and IP properties, and well-defined with respect to both the approximation and exact Landau-Ginzburg formula) were generated for a sum of weights $w_{tot} \leq 200$, for systems of 7-weights producing Calabi-Yau five-folds.
Additionally, where the exact Landau-Ginzburg formula computation time was infeasible for systems of 8-weights, a complementary dataset of candidate transverse weight systems (satisfying the necessary intradivisible and IP properties, and well-defined with respect to just the approximation) was generated, again for a sum of weights $w_{tot} \leq 200$, leading to candidate Calabi-Yau six-folds.
Some preliminary analysis of this data, and the respectively computed topological properties is provided, with a thorough analysis, and its full generation for $w_{tot} > 200$ left for future work.

These datasets, the respective code for analysis and ML, and an example notebook illustrating functionality to check intradivisibility, compute Euler number, and compute exact and approximated Hodge numbers of an input weight system of any size, are all available at this work's respective \href{https://github.com/Tancredi-Schettini-Gherardini/P5CY4ML}{GitHub} repository.

%%%%%%%%%%%%%%%%%%%%%%%%%%%%%%%%%%%%%%%%%%%%%%%%%
\section*{Acknowledgement}
EH acknowledges support from Pierre Andurand over the course of this research.
TSG was supported by the Science and Technology Facilities Council (STFC) Consolidated Grants ST/T000686/1 “Amplitudes, Strings \& Duality” and ST/X00063X/1 “Amplitudes, Strings \& Duality”. New data were generated and analysed during this study.

%%%%%%%%%%%%%%%%%%%%%%%%%%%%%%%%%%%%%%%%%%%%%%%%%
%\appendix
%\section*{Appendix}

\addcontentsline{toc}{section}{References}
\bibliographystyle{utphys}
\bibliography{biblio}{}

\end{document}